\documentclass[sigconf]{acmart}

\AtBeginDocument{%
 }

\setcopyright{rightsretained}
\acmJournal{PACMMOD}
\acmYear{2024} \acmVolume{2} \acmNumber{1 (SIGMOD)} \acmArticle{37} \acmMonth{2} \acmDOI{10.1145/3639292}
\makeatletter
\gdef\@copyrightpermission{
  \begin{minipage}{0.2\columnwidth}
   \href{https://creativecommons.org/licenses/by/4.0/}{\includegraphics[width=0.90\textwidth]{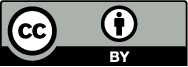}}
  \end{minipage}\hfill
  \begin{minipage}{0.8\columnwidth}
   \href{https://creativecommons.org/licenses/by/4.0/}{This work is licensed under a Creative Commons Attribution International 4.0 License.}
  \end{minipage}
  \vspace{5pt}
}
\makeatother

\usepackage{graphicx}
\usepackage[linesnumbered,ruled,vlined]{algorithm2e}
\usepackage{balance}
\usepackage{multirow}
\usepackage{enumitem}
\usepackage{soul}
\usepackage{tikz}
\usepackage{calc}
\usepackage{makecell}
\usepackage{diagbox}
\usepackage{pifont}
\usepackage{xcolor}
\usepackage{amsthm}  % Import the amsthm package for theorem-like environments
\newtheorem{lemma}{Lemma}  % Create a new lemma environment
\usepackage{framed}

\definecolor{shadecolor}{rgb}{0.92,0.92,0.92}

\newcommand*\circled[1]{\tikz[baseline=(char.base)]{
            \node[shape=circle,fill,inner sep=0.8pt] (char) {\textcolor{white}{#1}};}}
\newcommand*\dcircled[1]{\tikz[baseline=(char.base)]{
            \node[shape=circle,fill,inner sep=-0.03pt] (char) {\textcolor{white}{#1}};}}

\definecolor{purple}{rgb}{0.6,0.0,0.6}
\usepackage{tcolorbox}

\newcommand{\add}[1]{\textcolor{black}{{#1}}}

\newcommand{\annotate}[1]{\textcolor{gray}{{#1}}\xspace}

\newcommand{\oursys}{$\mathtt{PACE}$\xspace}

\newtheorem{case}{Case}

\begin{document}
\pagestyle{plain}

\title{PACE: Poisoning Attacks on Learned Cardinality Estimation}

\author{Jintao Zhang}
\orcid{0009-0001-6114-9429}
\affiliation{%
  \institution{Tsinghua University}
  % \streetaddress{1 Th{\o}rv{\"a}ld Circle}
  % \city{Hekla}
  \country{China}}
\email{zjt21@mails.tsinghua.edu.cn}

\author{Chao Zhang}
\orcid{0000-0002-8924-7629}
\authornote{Chao Zhang and Guoliang Li are the corresponding authors.}
\email{cycchao@tsinghua.edu.cn}
\affiliation{\institution{Tsinghua University}\country{China}}

\author{Guoliang Li}
\orcid{0000-0002-1398-0621}
\authornotemark[1]

\email{liguoliang@tsinghua.edu.cn}
\affiliation{\institution{Tsinghua University}\country{China}}

\author{Chengliang Chai}
\orcid{0000-0001-8080-5594}
\email{ccl@bit.edu.cn}
\affiliation{\institution{Beijing Institute of Technology}\country{China}}

\begin{abstract}
Cardinality estimation (CE) plays a crucial role in database optimizer. We have witnessed the emergence of numerous learned CE models recently which can outperform traditional methods such as histograms and samplings. However, learned models also bring many security risks. For example, a query-driven learned CE model learns a query-to-cardinality mapping based on the historical workload. Such a learned model could be attacked by poisoning queries, which are crafted by malicious attackers and woven into the historical workload, leading to performance degradation of CE. 

In this paper, we explore the potential security risks in learned CE and study a new problem of poisoning attacks on learned CE in a black-box setting.  There are three challenges. First, the interior details of the CE model are hidden in the black-box setting, making it difficult to attack the model.  Second, the attacked CE model's parameters will be updated with the poisoning queries, i.e., a variable varying with the optimization variable, so the problem cannot be modeled as a univariate optimization problem and thus is hard to solve by an efficient algorithm.  Third, to make an imperceptible attack, it requires to generate poisoning queries that follow a similar distribution to historical workload. We propose a poisoning attack system, \oursys, to address these challenges.  To tackle the first challenge, we propose a method of speculating and training a surrogate model, which transforms the black-box attack into a near-white-box attack. 
To address the second challenge, we model the poisoning problem as a bivariate optimization problem, and design an effective and efficient algorithm to solve it. 
To overcome the third challenge, we propose an adversarial approach to train a poisoning query generator alongside an anomaly detector, ensuring that the poisoning queries follow similar distribution to historical workload. 
Experiments show that \oursys reduces the accuracy of the learned CE models by 178$\times$, leading to a 10$\times$ decrease in the end-to-end performance of the target database. 
\end{abstract}




\received{July 2023}
\received[revised]{October 2023}
\received[accepted]{November 2023}

\maketitle

%!TEX root = ../main.tex

\section{Introduction} \label{sec:intro}

\noindent \textbf{Learned cardinality estimation.} 
Cardinality estimator is a vital component of the database query optimizer. In recent years, learned cardinality estimation (CE) methods~\cite{cidr2019/mscn, pvldb/DuttWNKNC19, fcnpool, vldb2020/deepdb, wu2020bayescard, pvldb/naru2019, pvldb/YangKLLDCS20, pvldb/FACE2022, guo2020multi, zhang2020selectivity, wang2023cardinality} have attracted significant attention due to their higher performance than traditional estimation methods such as histograms and sampling. 
However, learning-based models incur the risks of being attacked as the training data could be poisoned to degrade the estimation performance. 
In this work, we take query-driven cardinality estimation models~\cite{cidr2019/mscn, pvldb/DuttWNKNC19, fcnpool, ortiz2019empirical}, which are trained by fitting a set of training queries to their true cardinalities, as examples to study how to attack learned CE models by crafting poisoning queries. We discuss the attacks on data-driven CE models in Section~\ref{sec:con}.

\noindent \textbf{Motivation.} 
Nowadays, learned query-driven CE models have been deployed in real commercial systems~\cite{li2022htap, zhang2019unibench, zhang2021holistic, /pvldb/CDB_tutorial}, such as Microsoft Scope~\cite{wu2018towards}, Amazon Redshift AutoWLM~\cite{saxena2023auto}, GaussDB (for openGauss)~\cite{GaussDB,li2021opengauss}. 
Normally, machine learning models in online systems likely update themselves for maintaining high accuracy when some new training data are arrived~\cite{liberty2020elastic,he2020incremental,wu2018towards}. Similarly, query-driven CE models update themselves incrementally with newly executed queries~\cite{wu2018towards,sun2021learned,are_ready_ce}. This mechanism presents an opportunity for malicious people to craft some poisoning queries to attack the CE model, degrading the performance of the query optimizer.
Unfortunately, existing studies primarily focus on improving the performance of CE models while neglecting their potential vulnerabilities to poisoning attacks. To the best of our knowledge, this is the first work that studies the poisoning attack on learned CE models. 

\begin{figure}[!t] 
    \setcounter{figure}{0}
    \centering 
    
    \includegraphics[width=0.479\textwidth]{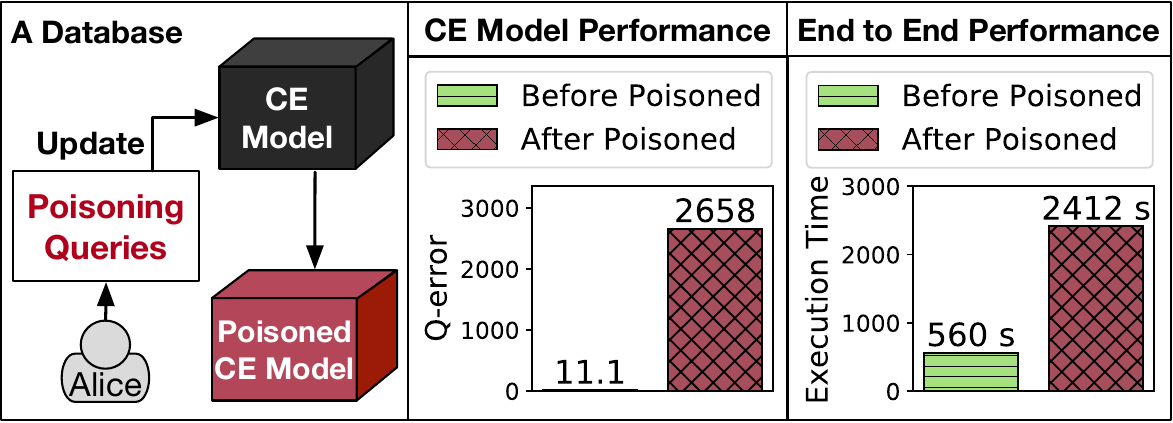}
    \vspace{-.6em}
        \caption{A example of a poisoning attack on a learned cardinality estimator.}
    \vspace{-1em}
    \label{fig:intro}
\end{figure}

\noindent\add{\textbf{Real scenarios of poisoning attacks in the context of databases.} Let us consider two motivating examples, an "internal" case (\texttt{Case}~\ref{_case1}) and an "external" case (\texttt{Case}~\ref{_case3}). The attackers in both cases have the incentives to poison a DB model.}

\add{
\begin{case}[Malicious Employee] \label{_case1}
    Suppose a scenario where an employee feels dissatisfied due to the unfair treatment or a notice of dismissal by his/her company. Since the company has a learning-based database used in production, s/he decides to perform a hidden act of retaliation. However, due to the company's strict permission policies, s/he has no deletion privilege and has only privilege of executing  SELECT SQL queries for operation and maintenance. To this end, s/he decides to attack the company's database with poisoning queries.
\end{case}
\begin{case}[Malicious Competitor] \label{_case3}
    Consider a situation where a cloud vendor wants to beat its competitor in order to get better reputation. This cloud vendor  deliberately rents a cloud database from its competitor and crafts malicious poisoning queries to attack the database's cardinality estimator in order to undermine the performance of the rented cloud database by poisoning attack. In this case, the cloud vendor not only has a strong incentive to carry out the attack but also has the authority to execute SQL queries on the target database.
\end{case}
}

\add{In addition, it's worth noting that previous research, such as that conducted by~\cite{kornaropoulos2022price}, has explored the issue of poisoning attacks on internal learned models in databases. Their study focused on poisoning attacks on learned indexes in a white box setting. In contrast, we study the poisoning attack in a black box setting (lack of all information about the target model), and there are more applications in realistic scenarios. In conclusion, we believe that the poisoning attacks on learned databases are a crucial topic that deserves more in-depth study.}

\noindent \textbf{Poisoning attack on learned CE.}  Considering a CE model used for estimating cardinalities of a given set of testing queries. 
The problem objective is to craft a small set of poisoning queries that, once used to update the CE model, would result in the model's lowest average estimation accuracy on the given test workload. As shown in Figure~\ref{fig:intro}, suppose Alice intends to attack the cardinality estimator in the database. She can craft some poisoning queries and execute them, causing the cardinality estimator in the database to be updated with these queries. As a result, for a same set of test queries, the average Q-error of the estimator increased from 11.1 to 2658, and the end-to-end execution time of the database increased from 560 seconds to 2412 seconds.

\noindent  \textbf{Challenges.} There are three challenges in poisoning attack on learned CE models in a black-box setting. 
First, the black-box setting of learned models prevent us learning how to generate poisoning queries by the updating direction of the CE model's interior parameters. 
Second, even under a white-box attack setting, efficiently solving the problem is difficult due to the updating of the CE model's parameters with the poisoning queries.
In other words, a parameter of the optimization objective is changing with the optimization variable. 
Third, there may be a significant divergence between the distribution of the poisoning queries and the historical workload, which could be easily detected~\cite{kurmanji2023detect}. Therefore, we need to generate queries that not only have poisoning effectiveness but also follow a similar distribution to the historical workload.

\noindent{\textbf{Our approach.}} To address these challenges, we propose a poisoning attack system \oursys that can attack learned query-driven cardinality estimators in a black-box setting. 
To address the first challenge, we propose a method for speculating the model type of the black box by comparing the similarities of the black-box model and candidate models' performance, followed by training a surrogate model based on the speculated model type. This enables us to convert the black-box attack into a near-white-box attack.  To address the second challenge, we propose to model the poisoning problem as a bivariate (i.e., poisoning queries and the CE model's parameters) optimization problem, and to achieve the optimization objective efficiently, we design an effective algorithm that utilizes a progressive update strategy to avoid unnecessary updates. 
To address the third challenge, we train an anomaly detector that can identify anomaly queries. We then employ an adversarial approach to train a poisoning query generator alongside the detector, ensuring that the distribution of the poisoning queries is similar to that of the historical workload.

\noindent  \textbf{Contributions:} We make the following contributions. 

\noindent (1) We study a new problem of poisoning attacks on learned cardinality estimation models in a black-box setting.

\noindent (2) We propose a method of speculating and training a surrogate model to transform black-box attack into a near-white-box attack. 

\noindent (3) We model the poisoning problem as a bivariate optimization problem, and design an algorithm that utilizes a progressive update strategy to achieve the optimization objective efficiently.

\noindent (4) We propose an adversarial approach to train a poisoning query generator alongside a trained anomaly detector, ensuring that poisoning queries follow a similar distribution to historical workload.

\noindent (5) We conducted extensive experiments, showing that our method reduces the accuracy of the learned CE models by 178$\times$, leading to a 10$\times$ decrease in the end-to-end performance of the database. And \oursys surpasses a basic algorithm by improving training efficiency by 9.7$\times$, and enhances the normality of poisoning queries by 72\%.

%% \end{itemize}

\section{Preliminaries}

\subsection{Query-driven Cardinality Estimation} \label{sec:2.1}

We focus on poisoning attacks over query-driven cardinality estimators~\cite{cidr2019/mscn, pvldb/DuttWNKNC19, fcnpool, ortiz2019empirical}. Given a set of queries $\mathbb{Q}=\{q_0, q_1, \cdots, q_n\}$ with their cardinalities $\mathbb{Y}=\{y_0, y_1, \cdots, y_n\}$, query-driven cardinality estimators will represent each query as a vector $x$. Then the training data of a query-driven cardinality estimator will be a set of pairs $(x, y)$. Finally, it learns a mapping from the query representations to the true cardinalities, which is regarded as a regression problem.

Formally, given a training workload  $\mathbb{D}_\text{train}=\{\mathbb{Q},\mathbb{Y}\}$, and a loss function $\mathcal{L}$, 
a query-driven CE model $f_w(x)$ is trained by an empirical risk minimization~\cite{vapnik1991principles} strategy, and finally the optimal parameter $w_b$ of CE model $f(\cdot)$ is obtained: 

\begin{equation} \setcounter{equation}{1}
    \label{equ:init_black_box}
    w_b \in \mathop{\arg\min}\limits_{w} \sum_{(x,y) \in \mathbb{D}_\text{train} }  \mathcal{L} (f_w(x) , y)
\end{equation}
where $\mathcal{L}$ is Q-error \cite{moerkotte2009preventing} loss, a most commonly used loss function in cardinality estimation. $\mathcal{L} (f_w(x) , y) = \frac{max(f_w(x), y)}{min(f_w(x), y)}$, where $f_w(x)$ is the estimated cardinality of a query and $y$ is the ground truth. $f_w(x)$ and $y$ are both greater than 0, because the last activation layer of the CE model limits the normalized value of $f_w(x)$ in $(0,1)$, and queries with $y=0$ will be eliminated during the training phase. This problem is usually solved by gradient descent~\cite{ruder2016overview}.

\begin{table}[!t] \setcounter{table}{0}
    \centering
    \caption{Notations.}
    \vspace{-.9em}
    \label{tab:notations}
    \setlength\tabcolsep{3pt}
    \scalebox{0.936}{
        \begin{tabular}{c|c||c|c} \toprule
        Notation  &  Description  &  Notation  &  Description \\ \hline
        $f_{w_b}(\cdot)$  &   Black-box model   &  $\mathcal{L}$   &   Loss function of CE model  \\ \hline
        $f_{s}(\cdot)$   &   Surrogate model   &  $\mathcal{L}_s$     &   Loss function of $f_{s}(\cdot)$  \\ \hline
        $f_{w_p}(\cdot)$  &   Poisoned model   &   $x$   &   Encoding of a query \\ \hline
        $\mathcal{D}$     &   Anomaly detector   &  $\mathcal{L}_d$     &   Loss function of $\mathcal{D}$  \\ \hline
        $\mathbb{D}_{\text{test}}$     &   Testing workload   &  $\mathbb{X}_p$   &   Poisoning queries  \\ \hline
        $\mathbb{Z}$     &   Gaussian noise   &  $\mathcal{G}$    &   Poisoning query generator  \\ \bottomrule
        \end{tabular}
    }
  
\end{table}

\subsection{\add{Threat Model}}  \label{sec:treat_model}
\add{\noindent \textbf{Adversary's goal:} The attacker's objective is to craft \textit{poisoning queries} $\mathbb{X}_p$ that can decrease the estimation accuracy of the target cardinality estimation model $f_{w}(\cdot)$ if it is updated using $\mathbb{X}_p$. The estimation accuracy refers to the Q-error~\cite{moerkotte2009preventing} of $f_{w}(\cdot)$ on a given test set $\mathbb{D}_\text{test}$, $\sum_{(x,y) \in \mathbb{D}_\text{test} }  Q\text{-}error (f_{w}(x), y)$. 
\\
\noindent \textbf{Adversary's knowledge:} We focus on \textit{black-box} attacks where the attackers cannot acquire the model type and specific parameters $w$ of the cardinality estimation model, and cannot get access to the data of the database and the training queries of the cardinality estimation model. The attackers can only obtain the database schema information to craft legal queries.
\\
\noindent \textbf{Adversary's capacity:} Attackers are able to get the true labels $\mathbb{Y}$ (i.e., cardinalities) of crafted queries by executing COUNT(*) SQLs and can inject poisoning queries as the training queries of the cardinality estimation model.
Moreover, attackers can obtain the estimated cardinalities $f_{w_b}(x)$ of the cardinality estimation model using the ``\textit{Explain}'' command.
\\
\noindent \textbf{Attack evaluation metrics.} We use four metrics as follows:
\textbf{(1)} \textit{Q-error}~\cite{moerkotte2009preventing} is a metric for evaluating the accuracy of a cardinality estimation model. \textbf{(2)} \textit{E2E latency} is used to quantify the end-to-end latency of query response in a database when utilizing a cardinality estimation model.
\textbf{(3)} \textit{Train\_Time} is used to evaluate the training time of the poisoning queries generation algorithm. 
\textbf{(4)} \textit{Divergence} is used to evaluate the normality of the poisoning queries distribution. Specifically, we use Jensen-Shannon Divergence~\cite{manning1999foundations} between the encodings of poisoning and historical queries. 
The higher the \textit{Q-error} and \textit{E2E latency} are, the more effective the attack is.
The lower the \textit{Train\_Time} of the algorithm and the \textit{Divergence} are, the more successful the attack is.}

\subsection{Problem Definition}

\noindent \textbf{{Poisoning Query Generation Problem.}} The studied problem can be formally defined as follows: Given a trained black-box CE model $f_{w_b}(\cdot)$, a testing workload $\mathbb{D}_{\text{test}}$. The studied problem is to craft a small poisoning workload $\mathbb{X}_p$ that can decrease the CE model's estimation accuracy if $f_{w_b}(\cdot)$ is updated to $f_{w_p}(\cdot)$ using $\mathbb{X}_p$. The objective is to maximize the estimation error of $f_{w_p}(\cdot)$ over $\mathbb{D}_{\text{test}}$:

\begin{equation}
\label{equ:bilevel opt} 
\mathbb{X}^*_p \in \mathop{\arg\max}\limits_{\mathbb{X}_p} \left( \mathcal{F}(\mathbb{X}_p) = \sum_{(x,y) \in \mathbb{D}_\text{test} }  \mathcal{L} (f_{w_p}(x), y) \right)
\end{equation}

In this work, we leverage the gradient information of the CE models with respect to the poisoning queries to carry out our attacks. This methodology supports attacking all query-driven CE models that are based on neural networks.

\subsection{Related Work} 

In the field of artificial intelligence security, many works~\cite{barreno2010security, nelson2008exploiting, xiao2012adversarial, zhang2017game, mei2015using, svm_p_attack, yang2017generative} attack machine learning models by tampering with the features or labels of the training data. These works typically adopt a white-box setting, meaning that the type and parameters of the victim model are known. In the field of databases, \cite{kornaropoulos2022price} studied the problem of a poisoning attack on learned indices in a white-box setting. However, in a real-world system, the learned model is essentially a black box to potential attackers, meaning that both the training data and the learned model itself are entirely inaccessible. Furthermore, even under a white-box setting, attacking the learned CE model remains challenging. First, existing works are unsuitable to attack learned CE models. For instance, \cite{kornaropoulos2022price} only considers the linear regression model. And most works~\cite{barreno2010security, nelson2008exploiting, xiao2012adversarial, zhang2017game, mei2015using, svm_p_attack, yang2017generative} can only produce poisoning samples of fixed dimension (e.g., fixed-size images). However, queries in learned CE models exhibit diverse join patterns that require specific treatment. Second, learned models in real-world scenarios are evolving, so it is necessary to craft poisoning queries efficiently. Otherwise, the attack could be obsolete. Third, to make an imperceptible attack, the poisoning queries should follow a similar distribution to the historical workload, otherwise it could be easily detected~\cite{kurmanji2023detect}.

%%$\mathcal{L}$ is a loss function determined by attacker. We use Q-error \cite{moerkotte2009preventing}, a most commonly used accuracy metric loss function in cardinality estimation~\cite{cidr2019/mscn, vldb2020/deepdb, pvldb/naru2019, pvldb/YangKLLDCS20, pvldb/FACE2022}. $\textit{Q-error} = \frac{max(\widehat{card}, card)}{min(\widehat{card},card)}$, where $\widehat{card}$ is estimated cardinality of a query and $card$ is the ground truth.

\begin{figure}[!t]
    \centering
\includegraphics[width=0.475\textwidth]{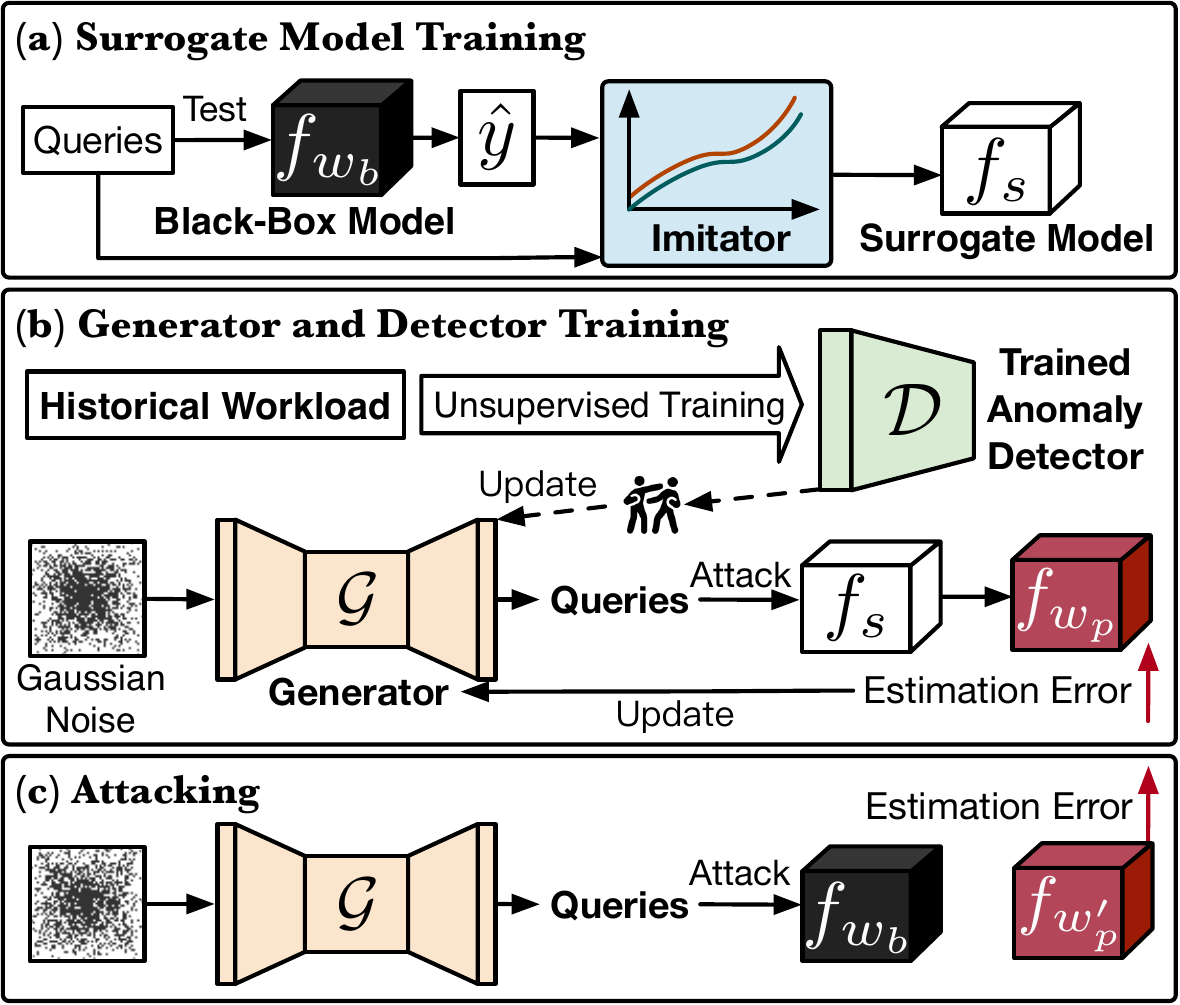}
    \vspace{-.6em}
    \caption{System overview. (\S 3)}
    \vspace{-.6em}
    \label{fig:Overview}
\end{figure}

\section{\oursys FrameWork}

We first provide an overview of \oursys in Section \ref{sec:overview}. Then, we describe the acquisition of surrogate CE model in Section \ref{subsec:surrogate} and poisoning query generation in Section \ref{subsec:generator and detector train}. Finally, we introduce how to use \oursys to attack the CE estimator in Section \ref{sec:attacking}.

\subsection{Overview} \label{sec:overview}

\noindent \textbf{{System Workflow.}} As shown in Figure \ref{fig:Overview}, the workflow of the \oursys can be summarized into three stages. 

\noindent \textit{\textbf{(a) Surrogate model acquisition.}} To cope with the black-box CE model, we propose to simulate the black-box model with a surrogate CE model $f_s(\cdot)$ by observing the input (queries) and output (estimated cardinalities) of the black-box model.

\noindent \textit{\textbf{(b) Poisoning Data Generator.}}
Given the surrogate model $f_s(\cdot)$ which is a white-box model to us, we treat the attack on $f_s(\cdot)$ as an attack on the original black-box model $f_{w_b}(\cdot)$ so that the parameter $w_b$ can be regarded as visible. To achieve the optimization goal in Equation~\ref{equ:bilevel opt}, we deploy a generator to generate poisoning queries. 
The basic idea is to train the generator $\mathcal{G}$ with the objective of maximizing the estimation error of the poisoned surrogate model $f_{w_p}(\cdot)$. 
At the same time, to avoid generating queries that are rather different from the historical workload, the generator will also fight against an anomaly detector that can detect abnormal queries compared to historical queries.

\noindent \textit{\textbf{(c) Attacking.}} The generated poisoning queries will be executed in the database, and the black-box CE model will be updated based on these queries, leading to larger estimation error.

\subsection{Surrogate CE Model Acquisition}  \label{subsec:surrogate}
The key for simulating a model is its type (CNN, RNN, etc.) and parameters. To derive a surrogate model $f_s(\cdot)$ according to the black-box $f_{w_b}(\cdot)$, we first speculate the type of $f_{w_b}(\cdot)$, followed by acquiring the parameters via training with the output from $f_{w_b}(\cdot)$.
Specifically, we train six CE models~\cite{cidr2019/mscn, pvldb/DuttWNKNC19, fcnpool, ortiz2019empirical} that contain all types of query-driven models based on neural networks (See subsection~\ref{subsec:exp_setting}). Next, we test these models as well as $f_{w_b}(\cdot)$ over some crafted queries. Finally, we select the model type that performs most similarly (including the accuracy and efficiency) to $f_{w_b}(\cdot)$. 
For the parameters of $f_s(\cdot)$, given the model type, an intuitive solution is to take the query encoding $x$ with the estimated result $f_{w_b}(x)$ of the black-box model as input, and trains a surrogate model according to a loss function like $\mathcal{L}(f_s(x),f_{w_b}(x))$
such that $f_s(\cdot)$ will be close to $f_{w_b}(\cdot)$. Unfortunately, solely relying on the output of the $f_{w_b}(\cdot)$ can lead to 
poor generalization performance on unseen queries.
To overcome this problem, we incorporate the ground-truth labels into the training process of model $f_{w_b}(\cdot)$, which can help $f_s(\cdot)$ to imitate $f_{w_b}(\cdot)$ better because $f_{w_b}(\cdot)$ also contains the information of these labels. Therefore, we propose to use both outputs of $f_{w_b}(\cdot)$ and the ground-truth labels as training examples. 
As a result,
the trained surrogate model $f_s^*(\cdot)$ has a better generalization performance in imitating $f_{w_b}(\cdot)$.
The details are introduced in Section \ref{sec:surrogate}. The experimental results (See Section \ref{sec:exp_surrogate}) indicate that, the parameters of the surrogate model are highly similar to that of the black box model after the simulation process, meaning that attacking $f_s^*(\cdot)$ is almost equivalent to attacking $f_{w_b}(\cdot)$.

\noindent \underline{\textbf{Remark.}} Our method can be easily extended to support new CE models. When a new CE model needs to be considered, we only need to expand the $k$ candidate models to $k+1$ candidate models.

\subsection{Generator and Detector Training}  \label{subsec:generator and detector train}
We propose a generation-based approach to produce the poisoning queries. To generate diverse queries, we take random Gaussian noise as input for the generator. 
In particular, the poisoning workload $\mathbb{X}_p$ can be generated by feeding Gaussian noise set $\mathbb{Z}=\{\vec z | \vec z \sim \mathcal{N}(0,1)$ \} to the generator $\mathcal{G}$ as follows. 
\begin{equation}
    \label{equ:generate_poison}
    \mathbb{X}_p = \mathcal{G}(\mathbb{Z})
\end{equation}

Consequently, our goal becomes how to train the generator $\mathcal{G}$ to generate poisoning queries that can maximize the loss of the poisoned CE model $f_{w_p}(\cdot)$.
Therefore, the optimization objective in Equation \ref{equ:bilevel opt} is as follows:
\begin{equation}
    \label{equ:opt_obj_G}
    \mathop{\arg\max}\limits_{\mathcal{G}} \ \mathcal{F}(\mathcal{G}(\mathbb{Z})) = \sum_{(x,y) \in \mathbb{D}_\text{test} }  \mathcal{L} (f_{w_p}(x), y)
\end{equation}

\noindent where $w_p$ is the surrogate model's parameters updated with $\mathbb{X}_p$.

We propose an efficient algorithm to iteratively train the generator following three steps: (1) We use the generator $\mathcal{G}$ to generate a number of poisoning queries $\mathbb{X}_p$. (2) $\mathbb{X}_p$ is temporarily used to update $f_s^*(\cdot)$ to obtain $w_p$.
(3) We update the generator to get closer to the optimization objective in Equation \ref{equ:opt_obj_G}. 
By repeating the above three steps,
$f_s^*(\cdot)$ is iteratively attacked, and thus the objective value $\mathcal{F}(\cdot)$ becomes larger. Finally, we stop training until the convergence $i.e.$, $\mathcal{F}(\cdot)$ is reached.
To ensure that the generated queries don't significantly deviate from the historical queries, we build a Variational Auto Encoder (VAE)~\cite{an2015variational} based anomaly detector $\mathcal{D}$ to counterbalance the generator.
Specifically, we use some historical queries in the database to train the anomaly detector according to a reconstruction loss $\mathcal{L}_d$. After training, a query will be deemed abnormal if its reconstruction error, as detected by $\mathcal{D}$, exceeds a certain threshold. 
Then, every time the poisoning queries are generated in the training phase, $\mathcal{D}$ will be triggered to detect abnormal queries among them. 
Finally, to prevent generating abnormal queries, the generator $\mathcal{G}$ is updated based on the reconstruction loss $\mathcal{L}_d$ associated with these abnormal queries. 
More details will be introduced in Section \ref{sec:nomality}.

\subsection{Attacking} \label{sec:attacking}

Given a batch of Gaussian noise $\mathbb{Z}$, the trained poisoning query generator $\mathcal{G}$ will output a batch of poisoning queries $\mathbb{X}_p$. 
Then we can run those queries in the target database. Afterward, the cardinality estimator $f_{w_b}(\cdot)$ in the database will use these queries and their true cardinalities to update itself. Eventually, the cardinality estimator could be poisoned and may not be able to accurately estimate the given testing workload.

\begin{figure*}[htbp]
	\centering
	\includegraphics[width=1.0\textwidth]{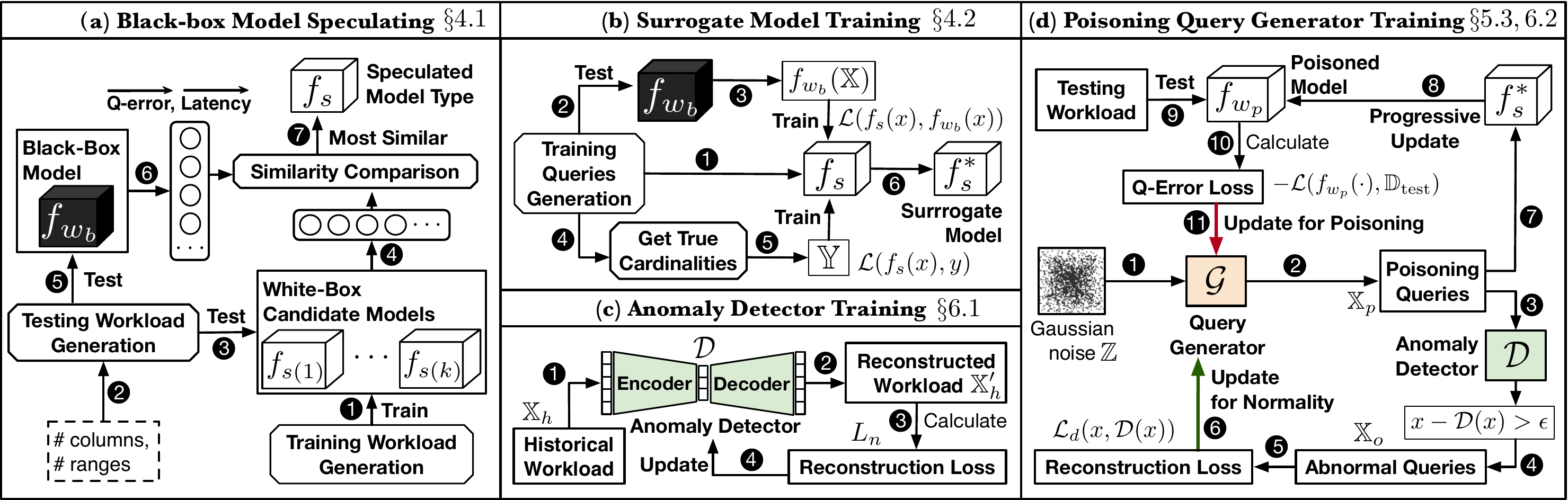}
    \vspace{-1.7em}
	\caption{Training workflow of PACE.}
 \vspace{-.5em}
	\label{fig:train_test}
\end{figure*}
%!TEX root=../main.tex

\section{Surrogate CE Model Acquisition} \label{sec:surrogate}

First, we introduce how to speculate the model type in Section \ref{sec:speculating}. Next, in Section \ref{sec:surrogate_training}, we introduce the strategy of training a white-box model to substitute the black-box model.

\subsection{Model Type Speculating of the Black Box} \label{sec:speculating}

\underline{\textbf{\textit{Key idea.}}} Before training a model, it is usually necessary to determine the model type. The reason is that the performance could be quite different even for a same task when using different model types. In our case, we aim to train a white-box model to replace a black-box model. To ensure that they perform similarly, it is essential for them to have the same model type. Therefore, the first step in acquiring a surrogate model is to speculate the type of the black-box model.
 We compare the performance similarity between the black-box model and the candidate models on a test workload with a specific distribution, and select the model type of the most similar candidate model.

To this end, we first assume $k$ different model types as candidates. Then we propose to pick one from the $k$ types as the type of $f_{w_b}(\cdot)$. At a high level, we pick the type comparing the performance of the $k$ models with that of $f_{w_b}(\cdot)$, based on a set $\mathbb{Q}$ of generated queries. Intuitively, given these queries, if one of the $k$ models performs the most similarly to $f_{w_b}(\cdot)$, they are likely to be with the same type. 
But note that queries in $\mathbb{Q}$ should have diverse properties ($e.g.,$ the column number), to test the performance variations across different model types. 
Because when queries have different column numbers and predicate range sizes, the estimated accuracy and latency are  much different on different model types. 
As proposed in \cite{are_ready_ce, sun2021learned}, (i) the accuracy of \texttt{MSCN} decreases less than \texttt{FCN} when the column number increases. (ii) the accuracy of \texttt{FCN} will be lower than other types when the range of filter predicates is too large or too small, (iii) the inference latency of \texttt{RNN} will increase as the number of columns increases. 
Specifically,  we first assume the $k$ models with different types as candidates, each of which is trained by a batch of randomly generated training queries.
Second, considering the diverse property discussed above, we generate $n_t$ test queries by varying the number of columns and the range size of filter predicates in queries.
Third, we test the $k$ candidate models $f_1(\cdot), \cdots, f_k(\cdot)$ and the black-box model $f_{w_b}(\cdot)$ over these queries, and then compute the mean of Q-error and estimation latency of the $k+1$ models for $n_t$ test queries, producing  $k+1$ vectors $\vec s_1, \cdots, \vec s_k, \vec s_b$ with $2 \times n_t$ dimensions.
Finally, we calculate the cosine similarity between $\vec s_1, \cdots, \vec s_k$ and $\vec s_b$, and pick the type with the highest similarity as the speculated type. 
That is, the speculated type is the same as the model type of $f_{i^*}(\cdot)$, where $i^*$ conforms to the following equation:

\begin{equation}
    \label{equ:speculate}
    i^* = \mathop{\arg\max}\limits_{i} \ Cosine(\vec s_i, \vec s_b) = \frac{\vec s_i \cdot \vec s_b}{||\vec s_i || \times ||\vec s_b||}
\end{equation}

\subsection{Training Strategy} \label{sec:surrogate_training}
\underline{\textbf{\textit{Key idea.}}} After determining the model type, the next step is to train the model parameters.  
The objective of this task is to enable the white-box model to perform as closely as possible to the black-box model, which requires both models to predict similar outputs for any given input. We utilize not only the estimated cardinalities of the black-box model, but also the ground-truth cardinalities as supervisory information in training the surrogate model. In this way, the surrogate model can achieve strong generalization performance in imitating the black-box model. 
A natural approach to training the surrogate model is to use the output of the black-box model as a supervisory information. This can be achieved by initializing the surrogate model with the speculated model type, and generating a batch of training queries $\mathbb{X}$.
The surrogate model $f_s(\cdot)$ can then be trained by imitating the output of the black-box model $f_{w_b}(x)$ on $\mathbb{X}$, where the training loss is defined as follows:
\begin{equation}
    \label{equ:intuitive_surrogate_train}
    \mathcal{L}_s(\mathbb{X}) = \sum_{x \in \mathbb{X} } \mathcal{L} (f_s(x), f_{w_b}(x))
\end{equation}

Considering that the black-box model $f_{w_b}(\cdot)$ is trained with true cardinalities in the database as supervision, so $f_{w_b}(\cdot)$ contains the information of true cardinalities. By incorporating the ground-truth labels $\mathbb{Y}$ of $\mathbb{X}$, the surrogate model can learn to better capture the underlying relationships between queries and their true cardinalities, leading to improved generalization performance in imitating $f_{w_b}(x)$. To this end, we propose a training method that can utilize the information not only from $f_{w_b}(x)$ but also $\mathbb{Y}$ and achieve a smaller imitation error. The loss function is as follows:
\begin{equation}
    \label{equ:our_surrogate_train}
    \mathcal{L}_s(\mathbb{X}, \mathbb{Y}) =  \sum_{x \in \mathbb{X}, y \in \mathbb{Y} } \left (\mathcal{L} (f_s(x), f_{w_b}(x)) + \mathcal{L} (f_s(x), y) \right )
\end{equation}

The former term $\mathcal{L} (f_s(x), f_{w_b}(x))$ can make $f_s(\cdot)$ imitate the $f_{w_b}(\cdot)$ well. And the latter item $\mathcal{L} (f_s(x), y)$ enables the $f_s(\cdot)$ generalizable for unseen queries.

\noindent \underline{\textbf{Remark.}} For the hyperparameters of the surrogate model, we design a default set of parameters, and we will analyze the impact of the inconsistency of hyperparameters in Section \ref{sec:exp}.

%!TEX root=../main.tex

\section{Poisoning Query Generation} \label{sec:generator}

We first present a high level idea of the training process of the poisoning query generator in Section~\ref{sec:5.1}. Afterward, we describe the query representation process and the structure of our poisoning query generator in Section \ref{sec:design_of_generator}.
Finally, we propose an efficient algorithm for training the poisoning query generator in Section~\ref{sec:algorithm}.

\subsection{High Level Idea} \label{sec:5.1}

In order to obtain the diverse poisoning queries, we employ a generator to learn the distribution of poisoning queries and subsequently generate them. As is common in generative networks~\cite{goodfellow2020generative, mirza2014conditional}, we provide the generator with Gaussian-distributed noise as input, enhancing its ability to output a variety of queries. Essentially, we enable the generator to learn and transform this Gaussian distribution into the distribution of poisoning queries. Two crucial aspects merit attention. Firstly, to enable the generation of poisoning queries with diverse join patterns, we design a join predicate generator, which creates valid join patterns, then pass them to the predicate generator as a part of input. Secondly, we train the generator using the estimation error of the attacked surrogate model, which serves as the overall objective function.

As shown in Figure~\ref{fig:train_test}(d), during each step of the training process of the generator, we feed Gaussian noise to the generator, which outputs poisoning queries (\circled{1}-\circled{2}).
These poisoning queries are used to update the surrogate model (\circled{7}-\circled{8}), and we use maximizing the estimation error of the updated surrogate model as the objective function (\circled{9}-\dcircled{11}). 
Since the whole process from the generation of poisoning queries by the generator to the updating of the white-box alternative model is derivable, we use a gradient descent method to update the generator.

\begin{figure*}[htbp]
  \centering
  \includegraphics[width=1.0\textwidth]{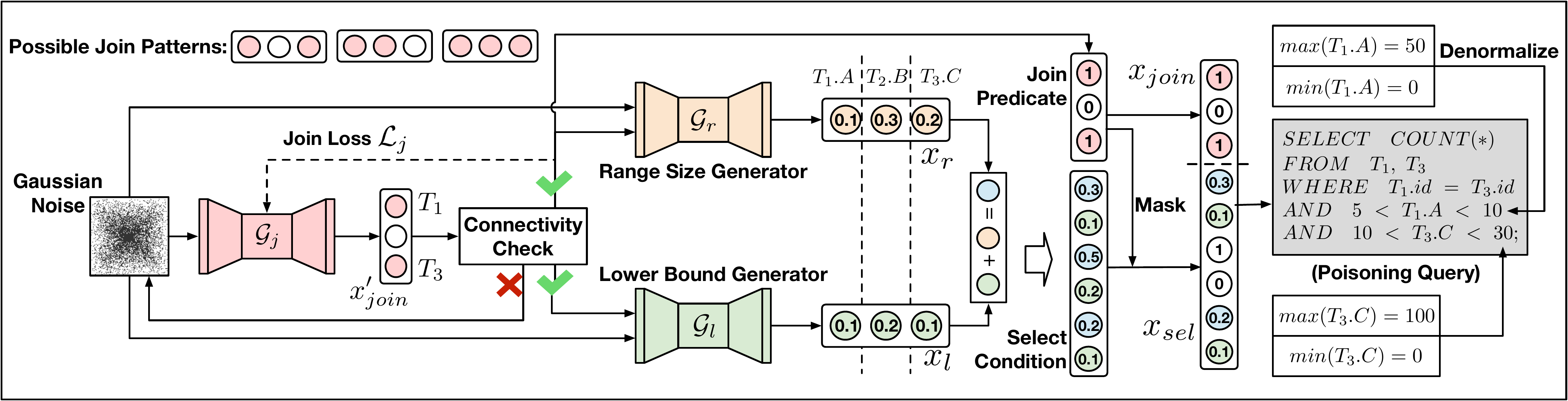}
  \vspace{-1.6em}
  \caption{Process of generating a poisoning query. (\S 5.2)}
  
  \label{fig:Design of generator}
\end{figure*}

\subsection{Generator Design}  \label{sec:design_of_generator}

For poisoning queries, we leverage a neural network-based generator to generate them.
In this part, we first introduce how we represent a query. After that, we give the detailed structure of the poisoning query generator.

\noindent \textbf{Query Representation.} 
Because most of the current learned CE methods only support SPJ queries, we focus on the generation of SPJ poisoning queries. Formally, consider a database with $n$ tables $\{T_1,..., T_i,..., T_n\}$ and $m$ attributes $\{A_1^1,..., A_i^j,..., A_n^m\}$. 
Since the cardinality of a SQL query $Q$ can be determined by two parts, namely the join predicate $J = \bowtie\{T_i\}$ and selection conditions $S = \sigma\{lb^j_i<A^j_i<ub^j_i\}$. Where $lb^j_i$ and $ub^j_i$ represent the normalized upper and lower bounds of the filtering predicate on the attribute $A^j_i$. 
The representation process for a query $Q=(J,S)$ into a vector $x$ involves several steps. 
First, the join predicate $J$ undergoes binary encoding to produce $n$-dimensional $x_{join}$, which consists of 0 and 1, where 1 means that the corresponding table lies in $J$, and 0 otherwise. 
Second, the selection condition $S$ is encoded into a vector $x_{sel}$ with a dimension of $2 \times m$, containing the normalized upper and lower bounds $(lb^1_1, ub^1_1, \cdots, lb^m_n, ub^m_n)$ of the filtering predicates corresponding to the $m$ attributes. In cases where $S$ does not contain a certain attribute $A^j_i$, the corresponding upper and lower bounds $[lb^j_i, ub^j_i]$ should be [0,1]. Finally, the representation result $x$ is obtained by concatenating $x_{join}$ and $x_{sel}$.

\noindent \textbf{Generator Structure.} To guarantee the diversity (i.e., considering various join combinations of tables) and correctness (i.e., ensuring that the upper bound is greater than the lower bound for each filter predicate) of the generated poisoning queries, we design a generator with three sub-generators based on deep neural networks.

Figure \ref{fig:Design of generator} depicts the process of generating a poisoning query, which is composed of 3 generators: the join predicate generator $\mathcal{G}_j$, the lower bound generator $\mathcal{G}_l$, and the range size generator $\mathcal{G}_r$.
The $\mathcal{G}_j$ sub-generator is responsible for generating various feasible join predicates for the poisoning queries to ensure the diversity. 
The $\mathcal{G}_l$ and $\mathcal{G}_r$ sub-generators are combined to generate the lower and upper bounds of the predicates of poisoning queries (the upper bound equals the lower bound plus the range size) according to the join predicates provided by $\mathcal{G}_j$ to ensure the correctness.

\noindent\textbf{Design of $\mathcal{G}_j$.} To ensure the diversity of the joins, we feed a Gaussian noise $z$ into  $\mathcal{G}_j$, and  it outputs a vector $x'_{join}$ of length $n$, indicating which tables are in the join predicate. 
The last layer of $\mathcal{G}_j$ is set as a sigmoid activation layer~\cite{han1995influence} to restrict the values in $x'_{join}$  between 0 and 1, and the value greater than 0.5 indicates that the corresponding table is in the join predicate, otherwise, it is not. 
After that, to ensure the correctness of the join predicates, we check whether the join predicate represented by $x'_{join}$ conforms to the join schema of the target dataset. If it is not satisfied, the Gaussian noise is regenerated and a new $x'_{join}$ is output. Otherwise, values greater than 0.5 in $x'_{join}$ are set to 1 and the rest are set to 0 to obtain the binary vector $x_{join}$. 
To  enhance the ability of  $\mathcal{G}_j$ to capture  correct join predicates, we construct a cross-entropy loss function $\mathcal{L}_j$ to train $\mathcal{G}_j$: 
\begin{equation}
\label{equ:join loss}
  \mathcal{L}_j(x_{join}', x_{join}) =  -\sum_{i=1}^n x_{join}'[i]\ log(x_{join}[i])
\end{equation}

\noindent\textbf{Design of ($\mathcal{G}_l$, $\mathcal{G}_r$).} 
To generate predicates, a combination of Gaussian noise $z$ and binary join vector $x_{join}$ is taken as the input of $\mathcal{G}_l$ and $\mathcal{G}_r$, allowing for the generation of diverse predicate upper and lower bounds with the specific join predicate. 
To guarantee the correctness of the generated predicates, the upper and lower bounds of the predicates for each attribute are not directly generated, as this could result in invalid queries with lower bounds greater than the upper bounds. 
Instead, $\mathcal{G}_l$ generates the lower bounds of the predicates, and $\mathcal{G}_r$ generates range size. The final layer of both $\mathcal{G}_l$ and $\mathcal{G}_r$ utilizes a sigmoid activation function to normalize the lower bounds and range size of the predicates between 0 and 1. Then we can ensure that the upper bounds are greater than the lower bounds because the upper bounds are calculated by adding the lower bounds and range size.
Then the $x_{sel}$ can be obtained by a masking process according to $x_{join}$. That is, if a table is not in the $x_{join}$, then the lower and upper bounds of the corresponding attributes are set to 0 and 1.

Finally, we can concatenate $x_{join}$ and $x_{sel}$ to obtain the representation $x$ of a poisoning query $Q$. And this representation can be easily transformed into a query according to the process of $x \rightarrow Q$.

\noindent \textbf{Summarization.} $\mathcal{G}_j$ transforms the Gaussian distribution into the correct join predicates. $\mathcal{G}_l$ and $\mathcal{G}_r$ transform the Gaussian distribution into selection predicates with poisoning effectiveness.

\subsection{Generator Training}  \label{sec:algorithm}

In this part, we discuss the methodology of training the generator. At a high level, the generator produces a batch of poisoning queries, which are utilized to update the surrogate model, and then use the estimation error of the updated model to guide the training of the generator ($i.e.,$ solving Equation~\ref{equ:opt_obj_G}). 
We first analyze the updating process of the surrogate model to determine its poisoned parameters. Next, we specify our objective function and propose a basic algorithm to train the generator. Due to the high time complexity of the algorithm, we finally propose an algorithm to improve efficiency.

\noindent \textbf{CE model updating.}  The surrogate model retains  existing parameters $w_b$ initially, updates itself on the poisoning queries for a small number ($K$) of iterations, and ultimately updates its parameters to $w^K_p$. The update process for one iteration can be formulated as:

\begin{equation}
    \label{equ:opt_st_}
    \begin{aligned}
    w^{e}_p = w^{e-1}_p - \alpha \nabla_{w^{e-1}_p} \sum_{(x,y) \in (\mathcal{G}(\mathbb{Z}),\mathbb{Y}_p) } \mathcal{L} \left( f_{w^{e-1}_p}(x), y \right), \ \ \  w^0_p = w_b
    \end{aligned}
\end{equation}

\noindent Where $w^e_p$ denotes parameter at the $e-$th iteration during the update process, and $e \in [0,K]$. $\alpha$ represents the learning rate. $w^0_p = w_b$ means the parameters of the black-box model are initially $w_b$.

\noindent \textbf{Objective function.} Once the parameters of the black-box model have been updated to $w^K_p$ after the $K$-step update process, our goal is to maximize  the objective function $\mathcal{F}$ by optimizing the parameters of the generator $\mathcal{G} = {\mathcal{G}_j, \mathcal{G}_r, \mathcal{G}_l}$:

\begin{equation}
    \label{equ:opt_obj_}
    \mathop{\arg\max}\limits_{\mathcal{G} = \{\mathcal{G}_j, \mathcal{G}_r, \mathcal{G}_l\}} \ \mathcal{F}(\mathcal{G}(\mathbb{Z}),\mathbb{Y}_p,w^K_p) = \sum_{(x,y) \in \mathbb{D}_\text{test} }  \mathcal{L} \left(f_{w^K_p}(x), y \right)
\end{equation}

Since $w^K_p$ varies with $\mathcal{G}$ as described in Equation~\ref{equ:opt_st_}, it is non-trivial to solve this optimization problem.

\add{\begin{lemma}[Bivariate optimization]
  The problem of poisoning query generation is a bivariate optimization problem that includes two variables, query generator $\mathcal{G}$ and poisoned model $w_p$. Particularly, $w_p$ is changing with $\mathcal{G}$ when maximizing the objective function.
\end{lemma}}

\noindent\add{\textbf{Analysis.} In the optimization objective as represented in Equation~\ref{equ:opt_obj_}, our goal is to maximize the function value $\mathcal{F}$ by optimizing $\mathcal{G}$. However, due to the updating process of the CE model on generated queries, $w_p$ is changing with $\mathcal{G}$ according to Equation~\ref{equ:opt_st_}. Therefore, the objective function must take into account the changing of $w_p$ when optimizing $\mathcal{G}$.}

\noindent\add{\textbf{Convergence analysis.} Generally, a non-convex optimization problem is guaranteed to converge only if its objective function has the property of Lipschitz continuous gradient~\cite{armijo1966minimization}. 
However, since this optimization objective includes a generative neural network and an update process of a CE model, the difficulty of expressing and deducing the large volume of parameters with mathematical formulas poses a notable challenge to prove the convergence from a mathematical perspective. 
In practice, such problems can typically be optimized to converge using the gradient descent method; namely, we can continually calculate the gradient of $\mathcal{G}$ for $\mathcal{F}$, and modulate $\mathcal{G}$ one step according to the gradient. 
In addition, to prevent the objective function from converging into a local optimum, we have used large steps in the case of small gradients for escaping from the local optimum. From the perspective of experimental verification, we will report the convergence curves of our optimization problem in Section~\ref{sec:exp_convergence}.}

\begin{figure}[!t]
	\centering
	\includegraphics[width=0.475\textwidth]{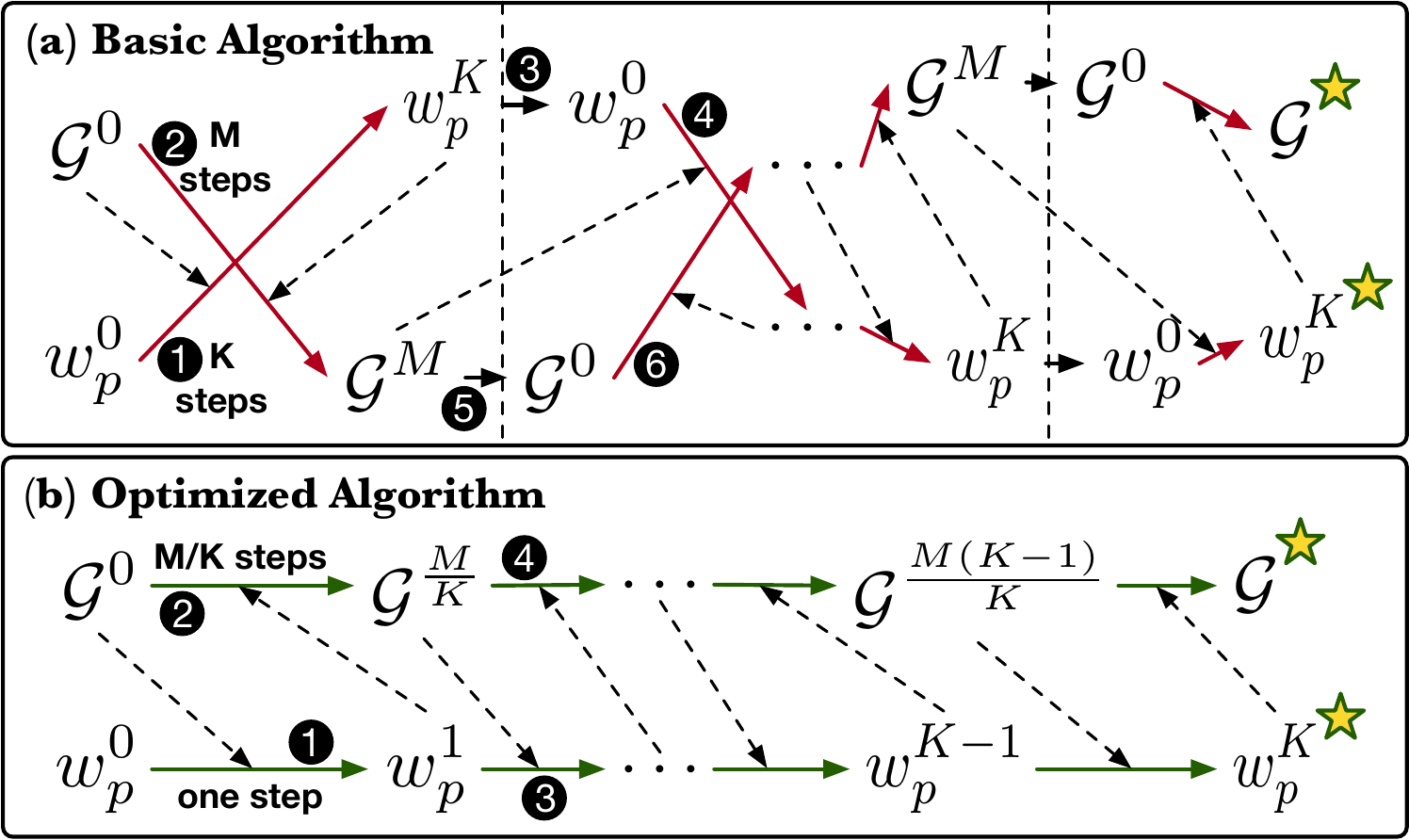}
        \vspace{-.6em}
	\caption{Analysis of the generator training. (\S 5.3)}
  \vspace{-.3em}
	\label{fig:analysis of algorithm}
\end{figure}

\noindent \textbf{Basic algorithm of solving equation~\ref{equ:opt_obj_}.} As shown in Figure \ref{fig:analysis of algorithm} (a), a feasible solution is proposed as follows:

\noindent\textbf{\underline{(1)}} Initialize $\mathcal{G}$ to $\mathcal{G}_0$, generate poisoning queries, and \circled{1} then update the parameters of the surrogate model to $w^K_p$.

\noindent\textbf{\underline{(2)}} Treat $w^K_p$ in the objective function as constants, and \circled{2} update $\mathcal{G}_0$ to $\mathcal{G}^M$ through the strategy of gradient descent until the objective function converges for the current $w^K_p$. We assume that this update process takes $M$ steps.

\noindent\textbf{\underline{(3)}} Obtain new poisoning queries by current $\mathcal{G}^M$. Then \circled{3} initialize $w_p$ as $w_p^0$, and \circled{4} obtain new $w^K_p$ by going through the process as shown in Equation \ref{equ:opt_st_} for $K$ steps. 

\noindent\textbf{\underline{(4)}} Repeat (2) and (3) until the objective function converges.

However, the shortcoming of this solution is that the algorithm complexity is  high. 

Assuming that (2) and (3) need to be repeated $n_r$ ($n_r>1$) times, then at each iteration of (2) and (3), $w_p$ is updated for $K$ steps under the current $\mathcal{G}$ instead of the optimal $\mathcal{G}$, 
and $\mathcal{G}$ is also updated by $M$ steps under the current $w^K_p$ instead of the poisoned $w^K_p$ by optimal $\mathcal{G}$. 
This leads to that a large number of updates of $\mathcal{G}$ and $w_p$ during each iteration are unnecessary. For example, in the beginning, the $w_p$ update updates $K$ steps for the initialized $\mathcal{G}$, and $\mathcal{G}$ updates $M$ steps according to current $w^K_p$, in which these $M$ steps are under the misleading guidance. 
In summary, the basic algorithm updates the generator and the surrogate model separately, which may result in unnecessary updates that do not improve the effectiveness of the poisoning queries.

\begin{algorithm}[!t]  

  \caption{Poisoning Query Generator Training \S 5.3,6.2}
  \label{alg:Generator Training} 
  \KwIn{Trained Surrogate CE Model $f_s^*(\cdot)$, trained Anomaly Detector $\mathcal{D}$, number of epochs for CE model updating $K$ and generator training $M$.}
  \KwOut{Trained poisoning query generator $\mathcal{G}$.}
  
  \For{o in [1, K]}{ 
      $\mathbb{Z}$ = $\mathcal{N}$(0,1); \tcp {\annotate{sample noise as generator input}}
      
      \For{i in [ 1+$\frac{M(o-1)}{K}$, $\frac{M \times o}{K}$ ]} {
          $\mathbb{X}_j' = \mathcal{G}_j(\mathbb{Z})$; \tcp {\annotate{join predicate vectors}}
          \If{Check($\mathbb{X}_j'$) == False} { 
            \tcp {\annotate{if join predicates not conform to schema}}
            Continue; \tcp {\annotate{regenerate $\mathbb{X}_j'$}}
          }
          $\mathbb{X}_j =$ Round($\mathbb{X}_j'$); \tcp {\annotate{round the values to 0 or 1}}
          $L_j = \sum_{(x_{join}' \in \mathbb{X}_j', x_{join} \in \mathbb{X}_j)} \mathcal{L}_j(x_{join}', x_{join})$\; % \tcp {loss for $\mathcal{G}_j$}
          $\mathcal{G}_j \leftarrow \mathcal{G}_j - \eta \nabla_{\mathcal{G}_j} L_j $\;
          $\mathbb{X}_s = \mathcal{G}_l(\mathbb{Z}, \mathbb{X}_j) + \mathcal{G}_r(\mathbb{Z}, \mathbb{X}_j)$; \tcp{\annotate{selection condition}}
          $\mathbb{X}_s =$ Mask$(\mathbb{X}_p, \mathbb{X}_j)$; \tcp {\annotate{mask $\mathbb{X}_p$ according to $\mathbb{X}_j$}}
          $\mathbb{X}_p =$ $\mathbb{X}_j \oplus \mathbb{X}_s$; \tcp {\annotate{poisoning queries}}
          $\mathbb{X}_o = \mathbb{X}_p [|\mathbb{X}_p - \mathcal{D}(\mathbb{X}_p)|>\epsilon]$; \tcp {\annotate{abnormal queries}}
          $L_n = \mathcal{L}_d (\mathbb{X}_o)$; \tcp {\annotate{reconstruction loss}}
          $\mathcal{G}_l \leftarrow \mathcal{G}_l - \eta \nabla_{\mathcal{G}_l} L_n $; \ \ \  $\mathcal{G}_r \leftarrow \mathcal{G}_r - \eta \nabla_{\mathcal{G}_r} L_n $\;
          $\mathbb{Y}_p$ = Query $(\mathbb{X}_p)$; \tcp {\annotate{get cardinalities of $\mathbb{X}_p$}}
          $f_\text{tmp}(\cdot) \leftarrow f_s^*(\cdot) - \alpha \nabla_{f_s^*(\cdot)} \mathcal{L} (f_s^*(\mathbb{X}_p), \mathbb{Y}_p) $\;
          $L_p = -\mathcal{L} (f_\text{tmp}(\cdot), \mathbb{D}_\text{test}$);  \tcp{\annotate{estimation error of temporarily updated surrogate model}}
          $\mathcal{G}_l \leftarrow \mathcal{G}_l - \eta \nabla_{\mathcal{G}_l} L_p $; \ \ \  $\mathcal{G}_r \leftarrow \mathcal{G}_r - \eta \nabla_{\mathcal{G}_r} L_p $\;
      }
      $f_s^*(\cdot) \leftarrow f_\text{tmp}(\cdot)$; \tcp {\annotate{update surrogate model}}
  }

  \textbf{return} $\mathcal{G}$\;
\end{algorithm}

\noindent \textbf{Acceleration algorithm.}
To overcome the shortcoming, we propose an efficient algorithm to train the poisoning query generator. 
At a high level, we can reduce unnecessary update steps by having $\mathcal{G}$ and $w_p$ interact in time. 
As shown in Figure \ref{fig:analysis of algorithm} (b) and Algorithm \ref{alg:Generator Training} (now we can ignore the anomaly detector $\mathcal{D}$ and lines 13-15, which will be introduced in Section~\ref{sec:nomality}), we can \circled{2} update the generator with a few steps after each \circled{1} update of the surrogate model. 
Specifically, for the inner loop shown in \textbf{lines 4-19}, we repeat the following processes (1-7) for $l/K$ times:

\noindent \textbf{\underline{(1)}} Input Gaussian noise $\mathbb{Z}$ to join predicate generator $\mathcal{G}_j$ to obtain the vector set $\mathbb{X}_j'$ and the binary vector set $\mathbb{X}_j$ of join predicates. 
 
\noindent \textbf{\underline{(2)}} Update the $\mathcal{G}_j$ according to the joining loss $L_j$. 
 
\noindent \textbf{\underline{(3)}} Input $\mathbb{Z}$ and $\mathbb{X}_j$ to lower bound generator $\mathcal{G}_l$ and range size generator $\mathcal{G}_r$ to obtain the selection vector set $\mathbb{X}_s$. 
 
\noindent \textbf{\underline{(4)}} Mask the predicates value of attributes that are not in the join predicates $\mathbb{X}_j$ in $\mathbb{X}_s$ to 0, and concatenate the $\mathbb{X}_j$ and $\mathbb{X}_s$ to get the vector set of generated poisoning queries $\mathbb{X}_p$. 

\noindent \textbf{\underline{(5)}} Obtain the cardinality labels $\mathbb{Y}$ of $\mathbb{X}_p$.

\noindent \textbf{\underline{(6)}} Update the surrogate model one step to $f_\text{tmp}(\cdot)$ on the poisoning queries (without updating the surrogate model itself). 
 
\noindent \textbf{\underline{(7)}} Take the estimation error of $f_\text{tmp}(\cdot)$ on $\mathbb{D}_\text{test}$ as the loss $L_p$, and update the generators $\mathcal{G}_l$ and $\mathcal{G}_r$ by one step according to $L_p$. 

And for each outer loop, we sample the Gaussian noise $\mathbb{Z}$ (\textbf{Line 2}) and assign the $f_\text{tmp}(\cdot)$ to the surrogate model (\textbf{Line 20}).

\add{\begin{lemma}[Algorithm Complexity]
  The time complexity of two generator training algorithms, the basic algorithm and acceleration algorithm, is $O(n_r*(M+K))$ and $O(M+K)$, respectively, where $M$ is the number of update steps required for the generator to converge, $K$ represents the update steps of the CE model on poisoning queries, and $n_r$ is the number of iterations to alternately update the generator and CE model so that the objective function converges. 
\end{lemma}}
  
  % \chao{introduce $n_r$, M, K, and compare the difference.}
\noindent\add{\textbf{Analysis.} For the basic algorithm, it contains $n_r$ update processes of the generator and the CE model, so its time complexity is $O(n_r*(M+K))$. The acceleration algorithm only contains one update process of the generator and the CE model, so its time complexity is $O(M+K)$.}

\noindent \textbf{Summarization.} We incrementally update the generator at each step of updating the surrogate model. This makes $\mathcal{G}$ and $w_p$ interact in time to reduce unnecessary update steps.

%!TEX root=../main.tex

\section{Ensure Distribution Consistency of Poisoning Queries} \label{sec:nomality}

If the distribution of generated poisoning queries are significantly different from the distribution of historical queries, the database may recognize these queries as abnormal and not use them to update the CE model. 
To address this problem, we train an anomaly detector using an unsupervised learning method. The anomaly detector is then deployed against the poisoning query generator.

\subsection{Anomaly Detector Training}  \label{sec:train anomaly detector}

\underline{\textbf{\textit{Key idea.}}} 
Typically, anomaly detectors are trained by labeling a batch of data as normal or abnormal, followed by training a classification model using supervised learning techniques. However, in our case, we do not have labeled normal and abnormal queries. Instead, we can obtain a set of historical queries $\mathbb{X}_h$, and if the generated poisoning queries distribution is similar to $\mathbb{X}_h$, these queries will not be considered as abnormal queries. 
So we can train an anomaly detector in an unsupervised way. During the training process, we use MSE loss to guide the anomaly detector to reconstruct the historical queries. Once completed, any query with a reconstruction error that surpasses a predetermined threshold is classified as abnormal.
Specifically, we utilize a variational auto-encoder (VAE)~\cite{an2015variational} as the anomaly detector $\mathcal{D}$ to reconstruct $\mathbb{X}_h$:

\begin{equation}
    \label{equ:variational autoencoder}
        x' = \mathcal{D}(x), \quad x \in \mathbb{X}_h 
\end{equation}

To train $\mathcal{D}$, we employ the Mean Squared Error (MSE) loss function~\cite{bickel2015mathematical} as the reconstruction loss.

\begin{equation}
    \label{equ:detector_loss}
        \mathcal{L}_d(\mathbb{X}_h) = \sum_{x \in \mathbb{X}_h } \frac{(x - x')^2}{ |\mathbb{X}_h| }
\end{equation}

Consequently, the VAE's ability to reconstruct a query depends on how different the query is from the distribution of queries in $\mathbb{X}_h$. The greater the consistency, the better the reconstruction performance of the VAE. 
After completing the training, we can assess whether an arbitrary query $x$ is abnormal or normal based on the reconstruction error $|x - x'|$. If the error is greater than a predetermined threshold $\epsilon$ (e.g., $\epsilon=0.1$ but $|x - x'|=0.15$), we classify the query as abnormal.

\subsection{Confrontation with Generator}  \label{deploy anomaly detector}
\underline{\textbf{\textit{Key idea.}}} 
We hope to adjust the parameters of the generator so that the poisoning queries $\mathbb{X}_p$ are not easily classified as abnormal queries by the anomaly detector. In essence, we strive to reduce the reconstruction error of the $\mathbb{X}_p$. 
$\mathbb{X}_p$ is generated by the generator, so the reconstruction loss of $\mathbb{X}_p$ can be backpropagated to the generator, so that we can update the parameters of the generator through the gradient descent method, thereby enhancing the ability of the generator to generate normal queries. 
During each training iteration of the poisoning query generator, we detect abnormal queries based on the anomaly detector and then use the reconstruction loss to update the generator. This ensures the normality of the $\mathbb{X}_p$ without significantly reducing the poisoning effect. 

As shown in Figure~\ref{fig:train_test}(c), in order to prevent the generator from generating abnormal queries, we can employ the anomaly detector $\mathcal{D}$ to identify abnormal queries $\mathbb{X}_a$ among the generated queries. Next, we can update the generator using the reconstruction loss of $\mathbb{X}_a$, which is $\mathcal{L}_d(\mathbb{X}_a)$.

Algorithm \ref{alg:Generator Training} outlines the process for updating the poisoning queries generator $\mathcal{G}$ using the reconstruction loss of abnormal queries in each inner loop. First, we select abnormal queries based on whether the reconstruction error of the generated queries $\mathbb{X}_p$ exceeds the threshold $\epsilon$ (\textbf{Line 13}). Then, we calculate the reconstruction loss $L_n$ of these abnormal queries (\textbf{Line 14}). Finally, we update $\mathcal{G}$ for one step by computing the gradient of the reconstruction loss with respect to the generator (\textbf{Line 15}). 

\noindent \add{\textbf{Mechanism analysis.} After completing the training process for the anomaly detector, as detailed in Section~\ref{sec:train anomaly detector}, the goal of generating normal queries is transformed into minimizing the reconstruction loss of generated queries (see Equation~\ref{equ:detector_loss}). To achieve the goal, we can get the reconstructed result $x'$ by inputting the generated query $x$ into the anomaly detector. Then, because the gradient between $x$ and $\mathcal{G}$ is differentiable, we can modulate the parameters of the generator $\mathcal{G}$ to minimize $(x-x')^2$ according to $\nabla_{\mathcal{G}} (x-x')^2$, i.e., keep updating $\mathcal{G}$ by a small step in the opposite direction of the gradient.}

It is important to note that utilizing the anomaly detector against the generator does not significantly reduce the poisoning effect. The rationale is that we are dealing with a dual-objective optimization problem, where we seek to maximize the effect of the poisoning queries while ensuring that their distribution similar to that of the historical queries. Corresponding to algorithm~\ref{alg:Generator Training}, the first update (\textbf{Line 15}) guarantees the normality of the generated queries, while the second update (\textbf{Line 19}) ensures the poisoning effectiveness.

\section{Experiments} \label{sec:exp}

This section evaluates the poisoning effect of \texttt{PACE}.
We mainly explore the following questions:

\begin{itemize} [leftmargin=*, align=parleft, labelsep=0.4em]
  \item \underline{(\S \ref{subsec:exp_on_ce_model})} Whether or not \oursys's attack on the cardinality estimation models is effective? And what are the differences among different types of cardinality estimation models?
  \item \underline{(\S \ref{subsec:exp_on_e2e})}
  What is the impact of the poisoning effect concerning the end-to-end query execution performance?
  \item \underline{(\S \ref{sec:exp_surrogate})}
  (1) Can \oursys accurately speculate the type of the black-box model? 
  (2) What is the impact if the model type of the surrogate model is different from that of the black-box model? 
  (3) How much better is our method that trains a surrogate model than directly training from the input and output of the black-box model?
  (4) What is the impact if the hyperparameters of the surrogate model and the black-box model are inconsistent? 
  (5) How does the number of poisoning queries $|\mathbb{X}_p|$ influence the effectiveness of the attack?
  \add{\item \underline{(\S \ref{sec:overhead})} What is the total overhead associated with \oursys, including training time, generation time, and attack time?}
  \item \underline{(\S \ref{sec:exp_efficiency})}
  How well does our poisoning query generation algorithm perform in terms of effectiveness and efficiency?
  \add{\item \underline{(\S \ref{sec:incremental})} How impactful are the poisoning queries when used on an incrementally trained CE model?}
  \item \underline{(\S \ref{sec:exp_normality})}
  Can the anomaly detector help generate the normality distribution while preserving the effectiveness of poisoning queries?\add{\item \underline{(\S \ref{sec:exp_convergence})} What is the real-world convergence performance of \oursys?}
\end{itemize}

\begin{figure*}[htbp]
  \centering
  \includegraphics[width=0.95\textwidth]{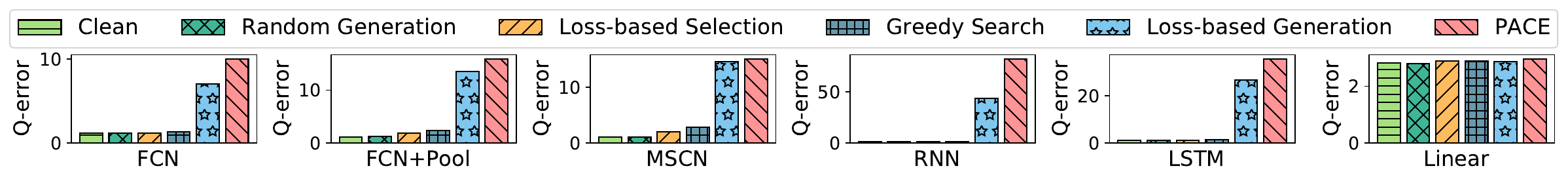}
  \vspace{-1.75 em}
  \caption{Mean of Q-error on DMV.}
  \vspace{-1.5 em}
  
  \label{fig:Qerror_DMV}
\end{figure*}

\begin{figure*}[htbp]
\centering
\includegraphics[width=0.95\textwidth]{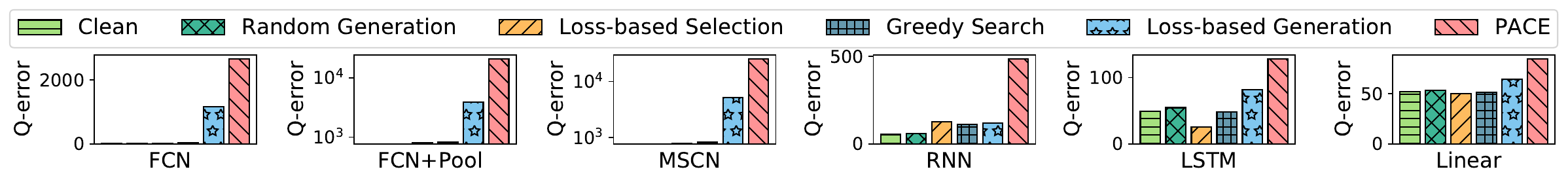}
 \vspace{-1.75 em}
\caption{Mean of Q-error on IMDB.}
  \vspace{-1.5 em}

\label{fig:Qerror_JOB}
\end{figure*} 

\begin{figure*}[htbp]
  \centering
  \includegraphics[width=0.95\textwidth]{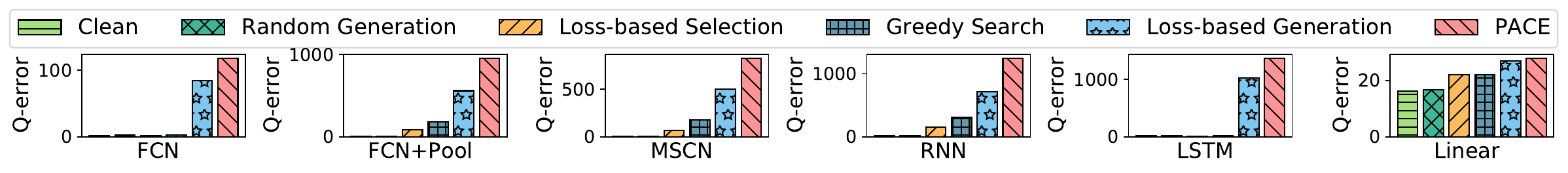}
   \vspace{-1.75 em}
  \caption{Mean of Q-error on TPC-H.}
  \vspace{-1.5 em}
  
  \label{fig:Qerror_TPCH}
\end{figure*}

\begin{figure*}[htbp]
  \centering
  \includegraphics[width=0.95\textwidth]{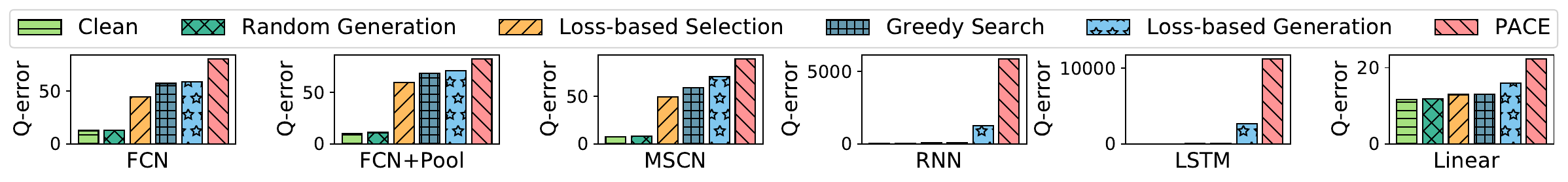}
    \vspace{-1.75 em}
  \caption{Mean of Q-error on STATS.}

  \label{fig:Qerror_STATS}
\end{figure*}

\begin{table}[h]
  \centering
  
  \caption{Query-driven CE models and their hyperparameters.}
  \vspace{-.9em}
  \label{tab:default model parameter}
  \setlength\tabcolsep{6pt}
  \scalebox{0.97}{
      \begin{tabular}{c||c|c|c|c} \toprule
      Model  &  \#Heads  &  \#Layers  &  HiddenDim  &  OutDim \\\hline
      \texttt{FCN}~\cite{pvldb/DuttWNKNC19}    &   1     &   4   &   $ 64 \times 128 $ &   1  \\\hline
      \texttt{FCN-Pool}~\cite{fcnpool}    &   1 $\times$ 3     &   4   &   $ 64 \times 128 $ &   1  \\\hline
      \texttt{MSCN}~\cite{cidr2019/mscn}     &   3 $\times$ 3     &   4   &   $ 64 \times 128 $ &   1  \\\hline
      \texttt{RNN}~\cite{ortiz2019empirical}      &   1     &   4   &   $ 64 $ &   1  \\\hline
      \texttt{LSTM}     &   1     &   4   &   $ 64 $ &   1  \\\hline
      \texttt{Linear} &   1     &   2   &   $ 128 $ &   1  \\ \bottomrule
      \end{tabular}
  }

\end{table}

\subsection{Experimental Setup} \label{subsec:exp_setting}

\textit{\textbf{\underline{Datasets.}}} 
We conduct experiments on 4 widely used datasets: \textbf{(1)} \texttt{DMV}~\cite{dmv} is a real-world single-table dataset that contains vehicle registration information in New York. 
\textbf{(2)} \texttt{IMDB}~\cite{leis2015good} is a movie rating dataset that consists of 21 tables.
\textbf{(3)} \texttt{TPC-H}~\cite{tpch} is a popular benchmark dataset that contains 8 tables. 
\textbf{(4)} \texttt{STATS}~\cite{vldb2022/alibaba_learnedEvaluation} dataset from the Stack Exchange network.
Since current query-driven cardinality estimation models fall short of learning string-type data, we encode the string-type attributes into numeric types using dictionaries.

\noindent \textit{\textbf{\underline{Workloads.}}}

For \texttt{DMV} and \texttt{TPC-H} datasets, we generate 10000 unseen training queries and 1000 testing queries similar to~\cite{li2022warper, pvldb/naru2019, pvldb/YangKLLDCS20} for each CE model. 
For \texttt{IMDB} and \texttt{STATS} datasets, we generate 10000 unseen training queries and 1000 testing queries based on the templates in \texttt{IMDB-JOB}~\cite{leis2015good} and \texttt{STATA-CEB}~\cite{vldb2022/alibaba_learnedEvaluation} respectively.

\noindent \textit{\textbf{\underline{CE models.}}} 
To verify the effectiveness of \oursys, we utilize all current neural network-based, query-driven Cardinality Estimation (CE) models, a total of six: 
\textbf{(1)} \texttt{FCN}~\cite{pvldb/DuttWNKNC19, fcnpool}. A lightweight fully connected neural network.  
\textbf{(2)} \texttt{FCN+Pool}~\cite{fcnpool}. A neural network that integrates 3 fully connected neural networks with a pooling layer. 
\textbf{(3)} \texttt{MSCN}~\cite{cidr2019/mscn}. A multi-set convolution network.  
\textbf{(4)} \texttt{RNN}~\cite{ortiz2019empirical}. A recurrent neural network.  
\textbf{(5)} \texttt{LSTM}. A long short-term memory network~\cite{sherstinsky2020fundamentals}. 
\textbf{(6)} \texttt{Linear}. A simple Linear regression network. 
The default hyperparameters of the 6 CE models as shown in Table~\ref{tab:default model parameter}. 

\noindent \textit{\textbf{\underline{Baselines.}}} 
We compare \texttt{\oursys} with the performance of the CE models before any attacks (\texttt{Clean}). We compare four baselines of crafting poisoning queries as follows:

\noindent \textbf{(1)} \texttt{Random Generation (Random)}. Randomly generate a set of queries as described in \textit{Workload} as poisoning queries. 

\noindent \textbf{(2)} \texttt{Loss-based Selection (Lb-S)}. Randomly generate a set of queries and select 10\% queries that maximize the inference loss $\mathcal{L}$ of the unpoisoned surrogate model. 

\noindent \textbf{(3)} \texttt{Greedy Search (Greedy)}. Randomly choose a join pattern from the possible join patterns. Then randomly generate 10 range select conditions for each attribute of all tables in the selected join pattern. Finally, construct a poisoning query by selecting one condition for each attribute, with the aim of maximizing the inference loss $\mathcal{L}$ of the unpoisoned surrogate model.

\noindent \textbf{(4)} \texttt{Loss-based Generation (Lb-G)}. Use the same generator as \oursys, but train with the goal of maximizing the inference loss $\mathcal{L}$ of the unpoisoned surrogate model.

\noindent \textit{\textbf{\underline{Hyper-parameters.}}} The default number of poisoning queries $|\mathbb{X}_p|$ is 450, which accounts for only 5\% of the training queries. The number of layers of $\mathcal{G}_j$, $\mathcal{G}_r$, $\mathcal{G}_l$ and $\mathcal{D}$ are 4, 5, 5, 7. The reconstruction threshold $\epsilon$ is 5\%. The learning rates $\alpha$ and $\eta$ are both $5e^{-3}$ and Adam \cite{kingma2014adam} optimizer is applied. 
The number of iterations $K$ for incremental updates of the CE model is 10.  
The number of iterations $l$ for generator training is 20. 
The number of outer loops of the basic algorithm $n_r$ is 20.

\begin{table*}[!t]
	\centering
	\caption{Percentile Q-errors Results.}
        \vspace{-.5em}
	\begin{minipage}{0.363\textwidth}
		\centerline{\textbf{(a)} DMV}
    \setlength\tabcolsep{3.65pt}
		\scalebox{0.933}{
		\begin{tabular}{c||c||c|c|c|c}
			\toprule
			CE Model  &  Method  &  90th  &  95th  &  99th  &  Max \\
			\midrule
			\multirow{5}{*}{\texttt{FCN}}  &  \texttt{Clean} & 1.263  & 1.321 & 1.388 & 1.420 \\
                                      &  \texttt{Random}  &  1.280  &  1.330 & 1.390 & 1.471  \\
                                      &  \texttt{Lb-S}  & 1.362  &  1.438 & 1.633 & 1.661  \\
                                      &  \texttt{Greedy}  & 	1.587  & 	1.667  & 	1.837  & 	2.190  \\
                                      &  \texttt{Lb-G}  &  23.37  &  25.55 & 28.01 & 31.63  \\
                                      &  \texttt{PACE}  &  \textbf{36.47}  &  \textbf{45.15} & \textbf{59.22} & \textbf{73.71} \\
			\hline
            \multirow{5}{*}{\makecell[c]{\texttt{FCN+}\\ \texttt{Pool}}}  &  \texttt{Clean} &  	1.182 &  	1.138 &  	1.347 &  	1.523  \\
                                                &  \texttt{Random}  & 	2.059 & 	1.867 & 	2.293 & 	2.233  \\
                                                &  \texttt{Lb-S}  & 	2.537 & 	2.81 & 	3.515 & 	3.975  \\
                                                &  \texttt{Greedy}  & 	3.287 & 	3.694 & 	4.939 & 	5.372  \\
                                                &  \texttt{Lb-G}  & 	33.67 & 	42.79 & 	59.15 & 	76.22  \\
                                                &  \texttt{PACE}  & 	\textbf{38.16}	 &  \textbf{46.57} & 	\textbf{63.60} & 	\textbf{88.22}  \\
			\hline
			\multirow{5}{*}{\texttt{MSCN}}  &  \texttt{Clean}  &  1.173 &  1.209 & 1.335 & 1.359 \\
                                      &  \texttt{Random}  & 1.214 &  1.253 & 1.315 & 1.407 \\
                                      &  \texttt{Lb-S}  & 2.643  &  2.782 & 3.402 & 3.646  \\
                                      &  \texttt{Greedy}  & 	3.686  & 	3.683  & 	4.773  & 	4.768  \\
                                      & \texttt{Lb-G}  &  37.75  &  46.61 & 63.74 & 81.02 \\
                                      &  \texttt{PACE}   &  \textbf{38.23}  &  \textbf{47.50} & \textbf{64.85} & \textbf{83.22} \\
			\hline
      \multirow{5}{*}{\texttt{RNN}}  &  \texttt{Clean}  &  1.212 &  1.247 & 1.328 & 1.352 \\
                                    &  \texttt{Random}  & 1.208 &  1.275 & 1.352 & 1.396 \\
                                    &  \texttt{Lb-S}  & 1.430  &  1.470 & 1.550 & 1.590  \\
                                    &  \texttt{Greedy}  & 	1.717  & 	1.699  & 	1.823  & 	1.863  \\
                                    & \texttt{Lb-G}  &  57.98  &  62.75 & 66.70 & 72.99 \\
                                    &  \texttt{PACE}  &  \textbf{107.8}  &  \textbf{117.0} & \textbf{130.1} & \textbf{145.9}   \\
			\bottomrule
		\end{tabular}
		}
	\end{minipage}
	\hfill
	\begin{minipage}{0.204\textwidth}
		\centerline{\textbf{(b)} IMDB}
    \setlength\tabcolsep{3.6pt}
		\scalebox{0.93}{
		\begin{tabular}{c|c|c|c}
			\toprule
			90th  &  95th  &  99th  &  Max \\
			\midrule
			22.37  &  51.31 & 159.6 & 247.2 \\
      25.68 &  42.91 & 165.5 & 321.5  \\
      33.25  &  61.99 & 106.4 & 202.8  \\
      72.60 & 	160.8 & 	279.1 & 	617.5  \\
      2792  &  5536 & $1.4e^4$ & $6.1e^4$ \\
      \textbf{4581}  &  $\textbf{1.4e}^\textbf{4}$ & $\textbf{4.1e}^\textbf{4}$ & $\textbf{1.3e}^\textbf{5}$  \\
			\hline
      25.73 & 	46.24 & 	123.7 & 	466.3  \\
      70.53 & 	144.7 & 	387.6 & 	1168.7  \\
      392.4 & 	1124 & 	1815 & 	3247  \\
      731.6 & 	2215 & 	3529 & 	7097  \\
      $1.5e^4$ & 	$3.5e^4$ & 	$8.5e^4$ & 	$3.2e^5$  \\
      $\textbf{2.2e}^\textbf{4}$ & 	$\textbf{6.2e}^\textbf{4}$ & 	$\textbf{2.3e}^\textbf{5}$ & 	$\textbf{5.7e}^\textbf{5}$  \\
            \hline
      12.30  &  31.41 &  184.2 &  204.9 \\
      70.09  & 148.7 &  335 & 1223 \\
      258.4  & 586.6  &  967.1 & 2612  \\
      635.4  &  	2190  &  	3431  &  	7555  \\
      $1.4e^4$ &  $3.2e^4$  &  $9.1e^4$ & $2.8e^5$  \\
      $\textbf{2.0e}^\textbf{4}$  &  $\textbf{6.0e}^\textbf{4}$  &  $\textbf{2.3e}^\textbf{5}$ & $\textbf{4.5e}^\textbf{5}$  \\
			\hline
      134.9  &  304.0 &  478.7 & 1632  \\
      110.4  & 356.0 &  555.6 & 1691 \\
      387.5  & 846.0  &  1434 & 1656  \\
      346.7  &  	600.7  &  	1217  &  	1349  \\
      201.6  &  387.8 &  793.3 & 8689 \\
      \textbf{754.1} &  \textbf{1664} &  \textbf{6377} & $\textbf{4.5e}^\textbf{4}$   \\
			\bottomrule
		\end{tabular}
		}
	\end{minipage}
	\hfill
	\begin{minipage}{0.204\textwidth}
		\centerline{\textbf{(c)} TPC-H}
        \setlength\tabcolsep{3.6pt}
		\scalebox{0.93}{
    \begin{tabular}{c|c|c|c}
      \toprule
      90th  &  95th  &  99th  &  Max \\
			\midrule
			3.699 &  5.787 & 10.60 & 13.11 \\
      3.374 & 6.560 &  24.47 & 45.92 \\
      3.685 & 5.887 & 13.04 &  23.64 \\
      5.300 & 	9.590 & 	24.55 & 	45.78  \\
      132.8 &  305.9& 1088 & 5363 \\
      \textbf{182.1} &	\textbf{352.8} &	\textbf{2883} &	\textbf{6127}  \\
			\hline
      2.450 & 	2.593 & 	7.995 & 	15.12  \\
      3.366 & 	4.239 & 	10.85 & 	22.88  \\
      163.6 & 	504.5 & 	1145 & 	2216  \\
      271.8 & 	845.1 & 	2146 & 	4914  \\
      221.4 & 	5244 & 	$1.4e^4$ & 	$1.5e^4$  \\
      \textbf{359.4} & 	\textbf{5351} & 	$\textbf{1.9e}^\textbf{4}$ & 	$\textbf{2.8e}^\textbf{4}$  \\
      \hline
      2.537	& 4.547	& 14.91	& 20.49 \\
      2.425 &	4.707	& 13.95	& 21.20 \\
      97.98	& 484.0	& 960.8	& 2257 \\
      248.3 & 	1931 & 	3895 & 	$1.0e^4$  \\
      161.3	& 4153	& $1.0e^4$	& $1.3e^4$  \\
      \textbf{485.2}	& \textbf{4087}	& $\textbf{1.8e}^\textbf{4}$	& $\textbf{2.7e}^\textbf{4}$ \\
			\hline
      37.12	 & 48.12	& 141.7 &	179.1  \\
      58.71	& 71.52	& 85.68	& 99.81 \\
      403.1	& 529.2	& 775.5	& 1038 \\
      687.1	& 	828.9	& 	1244	& 	1737  \\
      675.6	& 5598 &	$1.4e^4$ &	$2.5e^4$\\
      \textbf{1079}	& \textbf{8437} & $\textbf{2.6e}^\textbf{4}$	& $\textbf{3.3e}^\textbf{4}$ \\
			\bottomrule
		\end{tabular}
		}
	\end{minipage}
  \hfill
	\begin{minipage}{0.204\textwidth}
		\centerline{\textbf{(d)} STATS}
        \setlength\tabcolsep{3.6pt}
		\scalebox{0.9306}{
    \begin{tabular}{c|c|c|c}
      \toprule
      90th  &  95th  &  99th  &  Max \\
			\midrule
			15.57  &  31.42  &  99.07  &  600.2 \\
      15.85  &  29.35 &  108.1  &  615.5 \\
      75.85  &  143.2  &  217.2  &  1682 \\
      77.10  &  155.1  &  233.2  &  1699 \\
      79.09  &  177.8  &  1032  &  1887 \\
      \textbf{105.8}  &  \textbf{222.5}  &  \textbf{1261}  &  \textbf{3342} \\
      \hline
      11.49  &  20.25  &  58.91  &  310.6 \\
      20.29  &  44.42  &  91.96  &  223.3 \\
      62.75  &  93.07  &  189.3  &  414.9 \\
      77.22  &  114.2  &  231.6  &  527.5 \\
      79.29  &  157.1  &  440.5  &  945.1 \\
      \textbf{90.08}  &  \textbf{172.1}  &  \textbf{871.9}  &  \textbf{2619} \\
      \hline
      24.51  &  43.10  &  109.1  &  155.8 \\
      29.56  &  46.43  &  118.3  &  135.1 \\
      193.2  &  259.8  &  427.5  &  1268 \\
      257.6  &  382.2  &  566.3  &  2713 \\
      2818   &  3252   &  7594   &  9190  \\
      \textbf{3622}  &  $\textbf{4.5e}^\textbf{4}$  &  $\textbf{1.3e}^\textbf{5}$  &  $\textbf{2.2e}^\textbf{5}$ \\
      \hline
      13.09   &  27.01   &  72.53   &  432.2 \\
      13.31   &  34.35   &  93.53   &  521.9 \\
      34.01   &  55.05   &  158.4   &  329.4 \\
      42.01   &  56.31   &  198.4   &  394.9 \\
      76.21   &  149.3   &  575.9   &  1156 \\
      \textbf{81.57}   &  \textbf{156.8}   &  \textbf{944.5}   &  \textbf{2889} \\
			\bottomrule
		\end{tabular}
		}
	\end{minipage}
\label{tab:Q-errors}

\end{table*}

\begin{table}[!t]
	\centering
	\caption{Percentile Q-errors results for LSTM and Linear regression CE models.}
 \vspace{-.5em}
  \hfill
	\begin{minipage}{0.261\textwidth}
    \centerline{\textbf{(a)} DMV}
    \setlength\tabcolsep{4.5pt}
    \scalebox{0.89}{
    \begin{tabular}{c||c||c|c}
    \toprule
    CE Model  &  Method  &  95th  &    Max \\
    \midrule
    \multirow{5}{*}{\texttt{LSTM}}  &  \texttt{Clean}  &  1.262 &  1.375  \\
                                    &  \texttt{Random}  &  1.337 &  1.527  \\
                                    &  \texttt{Lb-S}  &   1.333 &  1.411 \\
                                    &  \texttt{Greedy}  &   1.496  &  1.819 \\
                                    & \texttt{Lb-G}  &  38.16 &  43.08 \\
                                    &   \texttt{PACE}  &  \textbf{52.73} &  \textbf{58.36}   \\
    \hline
    \multirow{5}{*}{\texttt{Linear}}  &  \texttt{Clean}  &  5.141 &  8.128  \\
                                    &  \texttt{Random}  &  5.131 &  8.046  \\
                                    &  \texttt{Lb-S}  &  5.528 &  8.724  \\
                                    &  \texttt{Greedy}  &   5.533 &  8.651 \\
                                    &  \texttt{Lb-G}   &  5.016 &  8.153  \\
                                    &   \texttt{PACE}  &  5.190 & \textbf{8.741}  \\
    \bottomrule
    \end{tabular}
    }
	\end{minipage}
	\hfill
	\begin{minipage}{0.103\textwidth}
    \centerline{\textbf{(b)} IMDB}
    \setlength\tabcolsep{4.5pt}
    \scalebox{0.89}{
    \begin{tabular}{c|c}
    \toprule
    95th  &  Max \\
    \midrule
    202.9 & 510  \\
    249.2 & 668.9  \\
    275.2 & 780.5  \\
    315.2 & 799.9  \\
    424.2 & 1168 \\
    \textbf{833.1} & \textbf{2275}  \\
	  \hline
    249.0 & 1269 \\
	  261.5 & 1581  \\
    271.2 & 1551 \\
    281.2 & 1873  \\
    291.3 & 2061 \\
    \textbf{495.0} & \textbf{2258} \\
	\bottomrule
	\end{tabular}
	}
	\end{minipage}
	\hfill
	\begin{minipage}{0.103\textwidth}
    \centerline{\textbf{(c)} TPC-H}
    \setlength\tabcolsep{4.5pt}
    \scalebox{0.89}{
    \begin{tabular}{c|c}
    \toprule
    95th  &  Max \\
	\midrule
  47.63	&  68.81 \\
  48.16	&  129.7 \\
  67.93 &	 212.1  \\
  77.14 &  204.9  \\
  8867	&  $3.2e^4$ \\
  $\textbf{1.1e}^\textbf{4}$	&  $\textbf{4.0e}^\textbf{4}$  \\
	\hline
  53.91	&  114.9 \\
	62.21	&  159.4  \\
  80.63	& 205.7 \\
  95.2 & 218.5  \\
  154.5	&  494.8 \\
  \textbf{180.1}	&  471.3 \\
	\bottomrule
	\end{tabular}
	}
 % \hfill
	\end{minipage}
\hfill
\label{tab:other-Q-errors}

\end{table}

\noindent \textit{\textbf{\underline{Metrics.}}} 
We use four metrics as described in Section~\ref{sec:treat_model}.

\noindent \textit{\textbf{\underline{Environment.}}} All experiments were performed on a server with a 20-core Intel(R) Xeon(R) 6242R 3.10GHz CPU, an Nvidia Geforce 3090ti GPU, and 256GB DDR4 RAM.

\subsection{Decline of the CE Models' Accuracy} \label{subsec:exp_on_ce_model}

\noindent \textbf{\underline{Average accuracy.}} The increase in the average Q-error can reflect the overall attack ability of the poisoning methods. 
Figure~\ref{fig:Qerror_DMV}, ~\ref{fig:Qerror_JOB}, ~\ref{fig:Qerror_TPCH} and ~\ref{fig:Qerror_STATS} show the average Q-error of each CE model before (\texttt{Clean}) and after being attacked on \texttt{DMV}, \texttt{IMDB}, \texttt{TPC-H} and \texttt{STATS} datasets. 
We can find that for \texttt{FCN}, \texttt{FCN+Pool}, \texttt{MSCN}, \texttt{RNN} and \texttt{LSTM}, the order of poisoning effectiveness is \oursys > \texttt{Lb-G} > \texttt{Greedy} > \texttt{Lb-S} > \texttt{Random} obviously. 
On average, \oursys outperforms the four baselines 2$\times$, 27$\times$, 55$\times$, 212$\times$ respectively. The reason for \oursys > \texttt{Lb-G} is that \texttt{Lb-G} only focuses on the inference loss before the model is poisoned, but the accuracy of the poisoned model is directly related to the inference loss after the model is poisoned. The reason for \texttt{Lb-G} > \texttt{Greedy} > \texttt{Lb-S} is that \texttt{Greedy} and \texttt{Lb-S} have no training process, resulting in a limited search space for poisoning queries. 
In addition, The attack effect of \oursys on \texttt{IMDB}, \texttt{TPC-H} and \texttt{STATS} is an order of magnitude higher than that on \texttt{DMV}. For example, on \texttt{IMDB}, \texttt{TPC-H} and \texttt{STATS}, \texttt{PACE} increases the estimation error by an average of 370$\times$, 138$\times$, and 89$\times$, respectively, while on \texttt{DMV} it is 26$\times$. This is because the simplicity of a single-table dataset makes it difficult to find queries that can significantly affect the CE models. 
And we find that the \texttt{FCN+Pool} and \texttt{MSCN} models perform very similarly on different datasets for each poisoning attack method. This is because the architectures of these two models are very similar~\cite{fcnpool}. 
Finally, for the \texttt{Linear} CE model, the effectiveness of all attack methods is not obvious, because the Linear regression model has few parameters, which reduces the fitting ability but improves the robustness of the model.

\noindent \textbf{\underline{Percentile accuracy.}} 
Percentile accuracy in cardinality estimation tasks is also important, especially high percentile accuracy, which can easily affect the performance of database query optimization~\cite{autoce, yu2022cost}. Table~\ref{tab:Q-errors} and ~\ref{tab:other-Q-errors} shows the percentile Q-error of each CE model before and after being attacked on different datasets. 

We can find that for the high percentile Q-error ($>$90-th), \oursys outperforms the \texttt{Lb-G}, \texttt{Greedy}, \texttt{Lb-S} and \texttt{Random} 2.4$\times$, 73$\times$, 135$\times$, 242$\times$ respectively, which is much larger than the average Q-error. This is because the optimization goal of \oursys is to find queries with the highest poisoning effectiveness directly.

\begin{table}[!t]
	\centering
 
	\caption{End-to-end execution time (s) results.}
  \vspace{-.6em}
  \setlength\tabcolsep{6pt}
		\scalebox{0.89}{
		\begin{tabular}{c||c||c|c|c|c|c}
			\toprule
			Dataset  &  Method  &  \texttt{FCN}  &  \texttt{FCN+Pool}  &  \texttt{MSCN}  &  \texttt{RNN}  &  \texttt{LSTM} \\
			\midrule
			\multirow{5}{*}{\texttt{IMDB}}  
          &  \texttt{Clean}   &  560.1  &  531.7  &  528  &  573.8  &  607.9 \\
          &  \texttt{Random}  &  586.4  &  533.2  &  528.1  &  600  &  632.3 \\
          &  \texttt{Lb-S}    &  657.5  &  712.5  &  682.6  &  792.7  &  691.6 \\
          &  \texttt{Greedy}  &  887.1  &  844.6  &  796.5  &  745.0  &  760.2 \\
          & \texttt{Lb-G}     &  1634  &  1740  &  2074  &  1072  &  918.1 \\
          &  \texttt{PACE}    &  \textbf{2412}  &  \textbf{3857}  &  \textbf{4656}  &  \textbf{1214}  &  \textbf{1160} \\
          \hline
          \multirow{5}{*}{\texttt{TPC-H}}  
          &  \texttt{Clean}   &  61.58  &  61.92  &  62.09  &  65.27  &  64.49 \\
          &  \texttt{Random}  &  62.36  &  64.68  &  65.36  &  68.70  &  69.10 \\
          &  \texttt{Lb-S}    &  87.26  &  223.2  &  234.9  &  96.64  &  149.7 \\
          &  \texttt{Greedy}  &  101.3  &  237.6  &  248.0  &  130.8  &  164.4 \\
          & \texttt{Lb-G}     &  140.5  &  262.6  &  258.7  &  193.9  &  198.6 \\
          &  \texttt{PACE}    &  \textbf{246.8}  &  \textbf{446.6}  &  \textbf{358.6}  &  \textbf{565.5}  &  \textbf{576.4} \\
          \hline
          \multirow{5}{*}{\texttt{STATS}} 
          &  \texttt{Clean}   &  24.26  &  23.78  &  23.47  &  24.28  &  24.19 \\
          &  \texttt{Random}  &  25.04  &  24.68  &  24.48  &  26.02  &  25.34 \\
          &  \texttt{Lb-S}    &  29.77  &  29.88  &  29.87  &  29.14  &  30.48 \\
          &  \texttt{Greedy}  &  36.80  &  43.64  &  39.61  &  38.70  &  39.45 \\
          & \texttt{Lb-G}     &  52.41  &  47.18  &  51.60  &  56.83  &  49.34 \\
          &  \texttt{PACE}    &  \textbf{190.1}  &  \textbf{180.1}  &  \textbf{185.9}  &  \textbf{282.4}  &  \textbf{247.1} \\
  \bottomrule
		\end{tabular}
		}
    \label{tab:E2E time}

\end{table}

\subsection{Impact on End-to-End Execution Time} \label{subsec:exp_on_e2e}
Decreased accuracy of the cardinality estimator often leads to degraded end-to-end execution performance of the database. In order to verify the effectiveness of \oursys on the end-to-end execution performance of the database, we compare \oursys and other baselines on the end-to-end execution time (\textit{E2E latency}) in the database. 

Table~\ref{tab:E2E time} shows the end-to-end execution time of 20 multi-table join testing queries using each CE model before (\texttt{Clean}) and after being attacked on \texttt{IMDB}, \texttt{TPC-H} and \texttt{STATS} datasets. 
We can find that for each dataset and CE model, \oursys achieves the longest end-to-end execution time. From the perspective of execution time increment, \oursys outperforms the \texttt{Lb-G}, \texttt{Greedy}, \texttt{Lb-S} and \texttt{Random} (2.5$\times$, 9.6$\times$, 14$\times$, 166$\times$), (2.6$\times$, 3.2$\times$, 3.7$\times$, 119$\times$) and (7$\times$, 24$\times$, 33$\times$, 173$\times$) on \texttt{IMDB}, \texttt{TPC-H} and \texttt{STATS} datasets respectively. 

This is because, for multi-table join queries, the accuracy of the cardinality estimation affects the join order and join operator selection of the query plan, both of which can lead to degradation of the end-to-end execution performance.

\subsection{Validation of Surrogate Model} \label{sec:exp_surrogate}

\begin{table}[!t]
  \centering
  \caption{Speculating accuracy for different CE model types.}
  \vspace{-.5em}
  \label{tab:speculating accuracy}
  \setlength\tabcolsep{4.5pt}
  \scalebox{0.93}{
      \begin{tabular}{c||c|c|c|c|c|c} \toprule
      \diagbox{Dataset}{Black-box}  &  \texttt{FCN}  &  \makecell[c]{\texttt{FCN+}\\ \texttt{Pool}}  &  \texttt{MSCN}  &  \texttt{RNN}  &  \texttt{LSTM}  & \texttt{Linear} \\ \hline
      \texttt{DMV}    &  75\%  &  75\%  &  75\%  &  85\%  &  80\%  &  95\%  \\ \hline
      \texttt{IMDB}   &  85\%  &  90\%  &  80\%  &  90\%  &  90\%  &  100\%  \\ \hline
      \texttt{TPC-H}  &  85\%  &  85\%  &  85\%  &  95\%  &  90\%  &  100\%  \\ \hline
      \texttt{STATS}  &  90\%  &  80\%  &  80\%  &  95\%  &  95\%  &  100\%  \\ \bottomrule
      \end{tabular}
  }

\end{table}

\begin{table}[htbp]
  
  \centering
  \caption{Decrease rates of attack effectiveness when the black-box model type is incorrectly speculated.}
  \vspace{-.5em}
  \label{tab:incorrect speculating}
  \setlength\tabcolsep{3pt}
  \scalebox{0.92}{
      \begin{tabular}{c||c|c|c|c|c|c} \toprule
      \diagbox{Surrogate}{Decrease}{Black-box}  &  \texttt{FCN}  &  \makecell[c]{\texttt{FCN+}\\ \texttt{Pool}}  &  \texttt{MSCN}  &  \texttt{RNN}  &  \texttt{LSTM}  & \texttt{Linear} \\ \hline
      \texttt{FCN}       &  0\%     &  1.74\%  &	5.6\%  &	6.75\%  &	3.85\%  &	3.35\%   \\\hline
      \texttt{FCN+Pool}  &  2.66\%  &  0\%  &	0.55\%  &	3.98\%  &	2.36\%  &	3.90\%   \\\hline
      \texttt{MSCN}      &  6.80\%  &  0.53\%  &	0\%  &	2.76\%  &	1.53\%  &	4.31\%  \\\hline
      \texttt{RNN}       &  29.2\%  &  8.80\%  &	7.68\%  &	0\%  &	2.86\%  &	2.00\%   \\\hline
      \texttt{LSTM}      &  13.1\%  &  4.14\%  &	1.59\%  &	2.34\%  &	0\%  &	6.36\%  \\\hline
      \texttt{Linear}  &  15.4\%  &  1.83\%  &	1.75\%  &	32.0\%  &	19.7\%  &	0\%  \\ \bottomrule
      \end{tabular}
  }

\end{table}

\noindent \textbf{\underline{Speculating accuracy.}} We randomly generate 20 sets of training queries on each dataset to train each type of CE model as a black-box model. Table~\ref{tab:speculating accuracy} shows the accuracy of speculating the type of black-box model using our speculating method. 
We can find that the average accuracy is 87.5\%, which illustrates the effectiveness of our speculating method. Among them, the accuracy rate of \texttt{FCN}, \texttt{FCN+Pool} and \texttt{MSCN} is the lowest, which is 82.1\% on average. This is because the architectures of the three models are so similar that they are easily speculated as one another. 

\begin{figure}[htbp]
  
	\centering
	\includegraphics[width=0.475\textwidth]{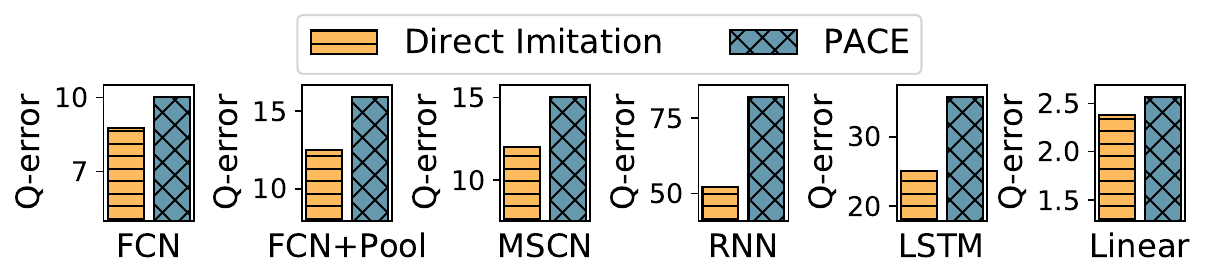}
 \vspace{-1em}
	\caption{Comparison of our imitation strategy and the direct imitation method.}
	\vspace{-.5em}
  \label{fig:imitate_strategy}
\end{figure}

\begin{figure}[htbp]
	\centering
	\includegraphics[width=0.49\textwidth]{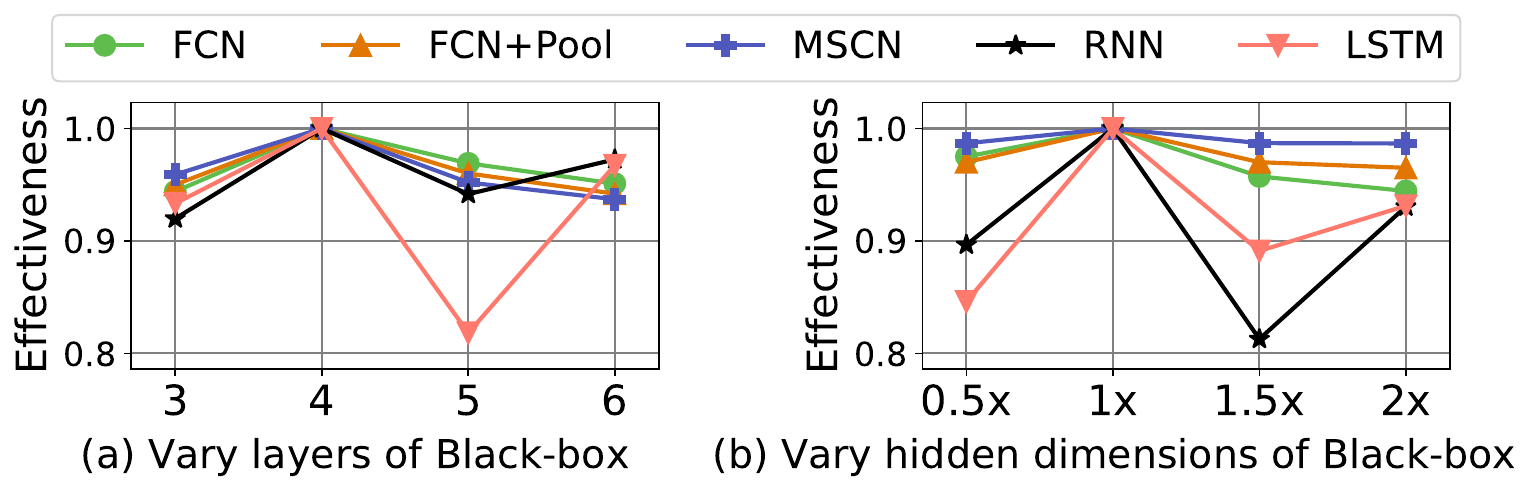}
  \vspace{-.9em}
	\caption{The Attack effectiveness when the hyperparameters of the black-box model change but the surrogate model maintains the default parameters.}
	
  \label{fig:exp_vary_parameters}
\end{figure}

\noindent \textbf{\underline{Incorrect speculation.}} To study the influence of the type of black-box model being speculated incorrectly on the attack effectiveness. We train a black-box model for each CE model on \texttt{DMV} dataset, and employ different types of models as surrogate models to study the rate of decline in attack effectiveness of \oursys. Table~\ref{tab:incorrect speculating} shows the decrease rates of attack effectiveness when the black-box model type is incorrectly speculated. 
We can find that the average decrease rate is 8.2\%, which indicates that overall even if the type of black-box model is incorrectly speculated, the decrease in the attack effectiveness is little.

\begin{table}[!t]
  \centering
  \caption{The multiplier by which Q-error increases when varying the number of poisoning queries.}
  \vspace{-.5em}
  \label{tab:vary N}
  \setlength\tabcolsep{5pt}
  \scalebox{0.95}{
      \begin{tabular}{c||c|c|c|c} \toprule
      \diagbox{Dataset}{Error Increase}{$|\mathbb{X}_p|$}  &  \texttt{225}  &  \makecell[c]{\texttt{450} \\ \texttt{(default)}}  &  \texttt{900}  &  \texttt{1800}  \\ \hline
      \texttt{DMV}   &   5.042$\times$    &  8.712$\times$  &  9.233$\times$   &  9.904$\times$   \\ \hline
      \texttt{IMDB}  &  130.9$\times$   &  241.6$\times$  &  262.3$\times$  &  265.9$\times$ \\ \bottomrule
      \end{tabular}
      }

\end{table}

\begin{figure}[htbp]
	\centering
	\includegraphics[width=0.485\textwidth]{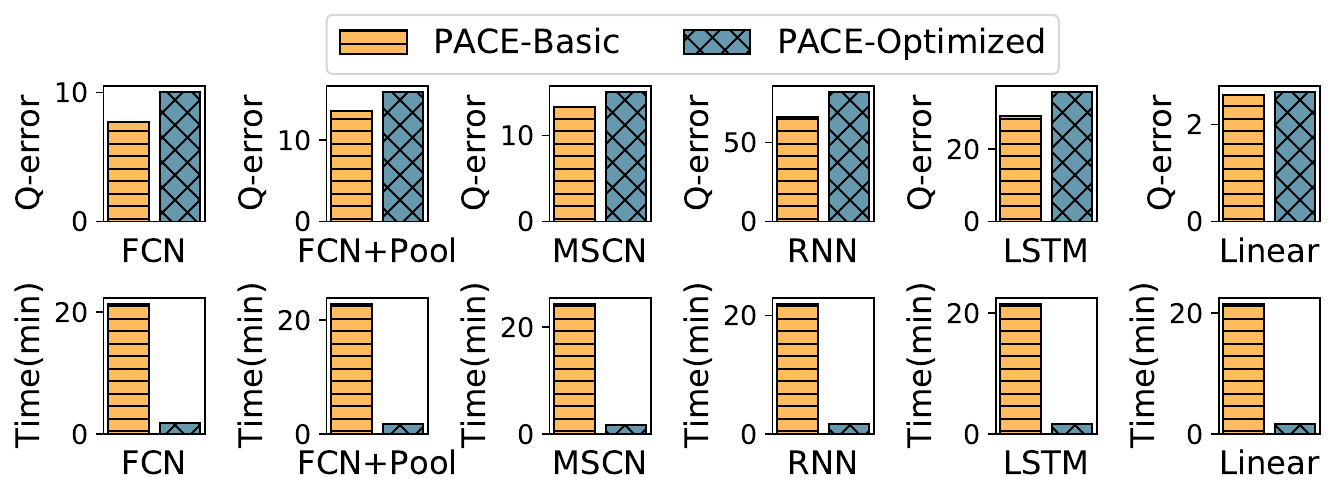}
        \vspace{-1em}
	\caption{Ablation of the algorithm optimization of PACE.}
  \vspace{-.5em}
	\label{fig:optimized_alg}
\end{figure}

\begin{figure}[htbp]
  
	\centering
	\includegraphics[width=0.485\textwidth]{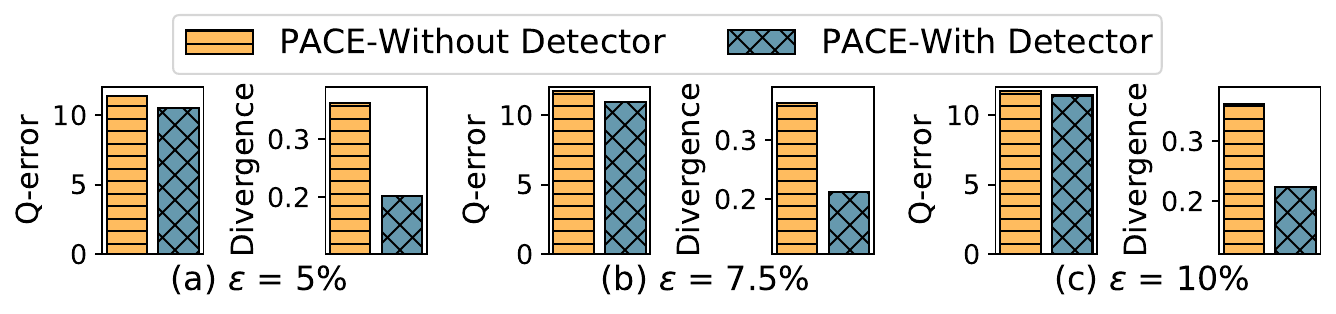}
  \vspace{-1em}
	\caption{Ablation of using the anomaly detector.}
  
	\label{fig:effective_of_detector}
\end{figure}

\noindent \textbf{\underline{Effectiveness of our training strategy.}}
To verify the superiority of our imitation strategy as in Equation~\ref{equ:our_surrogate_train} over the imitation method as in Equation~\ref{equ:intuitive_surrogate_train} (\texttt{Direct Imitation}), we compare the attack effectiveness of \oursys and using \texttt{Direct Imitation} on \texttt{DMV} dataset. 
As shown in Figure~\ref{fig:imitate_strategy}, we can find that the attack effectiveness of \oursys is on average 32.3\% higher than using \texttt{Direct Imitation}, which indicates that our imitation strategy is more effective. 

\noindent \textbf{\underline{Inconsistent hyperparameters.}} 
To study the impact of the inconsistency of hyperparameters of the black-box model and the surrogate model. We train black-box models with different layers and hidden dimensions for each CE model on \texttt{IMDB} dataset, and maintain the default hyperparameters shown in Table~\ref{tab:default model parameter} for the surrogate model. Figure~\ref{fig:exp_vary_parameters} shows the attack effectiveness of \oursys with the change of the black-box model hyperparameters. Where 1.0 on the vertical axis indicates the effectiveness when the hyperparameters of the surrogate model are the same as the black-box model. The horizontal axis in Figure~\ref{fig:exp_vary_parameters}(a) represents the number of layers of the black-box model. The horizontal axis in Figure~\ref{fig:exp_vary_parameters}(b) indicates the scale of hidden layers' dimension of the black-box model compared to the default hyperparameter. 
We can find that the average reduction rates are 5.5\% and 6.5\% for hyperparameters layers and hidden dimensions. This illustrates that the inconsistency of hyperparameters between the surrogate model and the black-box model has a small effect in general.

\noindent \textbf{\underline{Varying the number of poisoning queries.}} 
We employ a varying number of poisoning queries $|\mathbb{X}_p|$, to attack the \texttt{FCN} model.
Table~\ref{tab:vary N} shows the multiples of Q-error increase, relative to the pre-attack model, under the different numbers of $|\mathbb{X}_p|$. 
We can find that the desired level of attack efficacy is reached with as few as 450 poisoning queries, which is merely 5\% of the original training queries. Additional poisoning queries introduced beyond this threshold do not contribute noticeably to the improvement of the attack effectiveness.

\begin{table}[htbp]
  \centering
  \caption{\add{Overhead evaluation of \oursys on different datasets.}}
  \vspace{-.5em}
  \label{tab:overhead}
  \setlength\tabcolsep{6pt}
  \scalebox{1.0}{
    \add{
      \begin{tabular}{c||c|c|c} \toprule
      \diagbox{Dataset}{Time (s)}  &  Training  &  Generation  &  Attacking  \\ \hline
      \texttt{DMV}   &   188.88  &  0.5103   &  1.6069   \\ \hline
      \texttt{IMDB}  &  1711.0   &  0.5343   &  1.6355   \\ \hline
      \texttt{TPCH}  &  823.38   &  0.5014   &  1.7063   \\ \hline
      \texttt{STATS} &  3719.8   &   0.5240  &  1.6232   \\ \bottomrule
      \end{tabular}
    }
  }
\end{table}

\begin{table}[htbp]
  \centering
  
  \caption{\add{Overhead evaluation of \oursys on different numbers of generated queries.}}
  \vspace{-.5em}
  \label{tab:numbers_overhead}
  \setlength\tabcolsep{6pt}
  \scalebox{1.0}{
    \add{
      \begin{tabular}{c||c|c|c} \toprule
      \diagbox{Number}{Time (s)}  &  Training  &  Generation  &  Attacking  \\ \hline
      225 queries  &  188.88  &   0.2700  &  0.861  \\ \hline
      450 queries  &   188.88    &  0.5103  &  1.6069   \\ \hline
      900 queries  &  188.88  &   1.017  &  3.124  \\ \bottomrule
      \end{tabular}
  }
  }

\end{table}

\subsection{\add{Overhead Evaluation}} \label{sec:overhead}
\add{We conduct an experiment to evaluate the overhead of poisoning attacks, including the training time of \oursys, the generation time of poisoning queries, and the attacking time, i.e., the updating time of the target cardinality estimation model. Table~\ref{tab:overhead} provides the experiment results of \oursys's overhead on \texttt{FCN} across four datasets. The results indicate that the training time of \oursys is shortest on the \texttt{DMV} dataset because training on a single-table dataset eliminates the need to train the join predicate generator. The generation time of 450 queries is between 0.5s and 0.55s, and the attacking time of 450 queries is between 1.5s and 1.75s. Table~\ref{tab:numbers_overhead} shows the overhead of \oursys's under different numbers of poisoning queries on \texttt{DMV}. We find that under different numbers of generated queries, the training time will not change, but the generation and attacking time will proportionally change with the generated queries' number. The reason is that the ratio of the number of generated queries to the batch size determines the generation and attacking time. In summary, the attacking overhead is small.}

\subsection{Ablation of the Efficiency Optimization} \label{sec:exp_efficiency}
In this section, we will compare \oursys before and after algorithm optimization \texttt{PACE-basic} (See Figure~\ref{fig:analysis of algorithm}(a)) and \texttt{PACE-optimized} (See Figure~\ref{fig:analysis of algorithm}(b)) from the perspectives of effectiveness and efficiency. 

We explore the effectiveness and efficiency of \texttt{PACE-basic} and \texttt{PACE-optimized} respectively, on the \texttt{DMV} dataset. 
As shown in Figure~\ref{fig:optimized_alg}, we find that on average, \texttt{PACE-optimized} is 20.6\% more effective than \texttt{PACE-basic} in terms of attack effectiveness and 9.7 $\times$ faster in efficiency. 
The reason for the improvement in effectiveness is that \texttt{PACE-basic} updates the generator and the surrogate model separately, which results in unnecessary updates that do not improve the effectiveness of the poisoning queries.

\begin{figure}[htbp]
	\centering
	\includegraphics[width=0.489\textwidth]{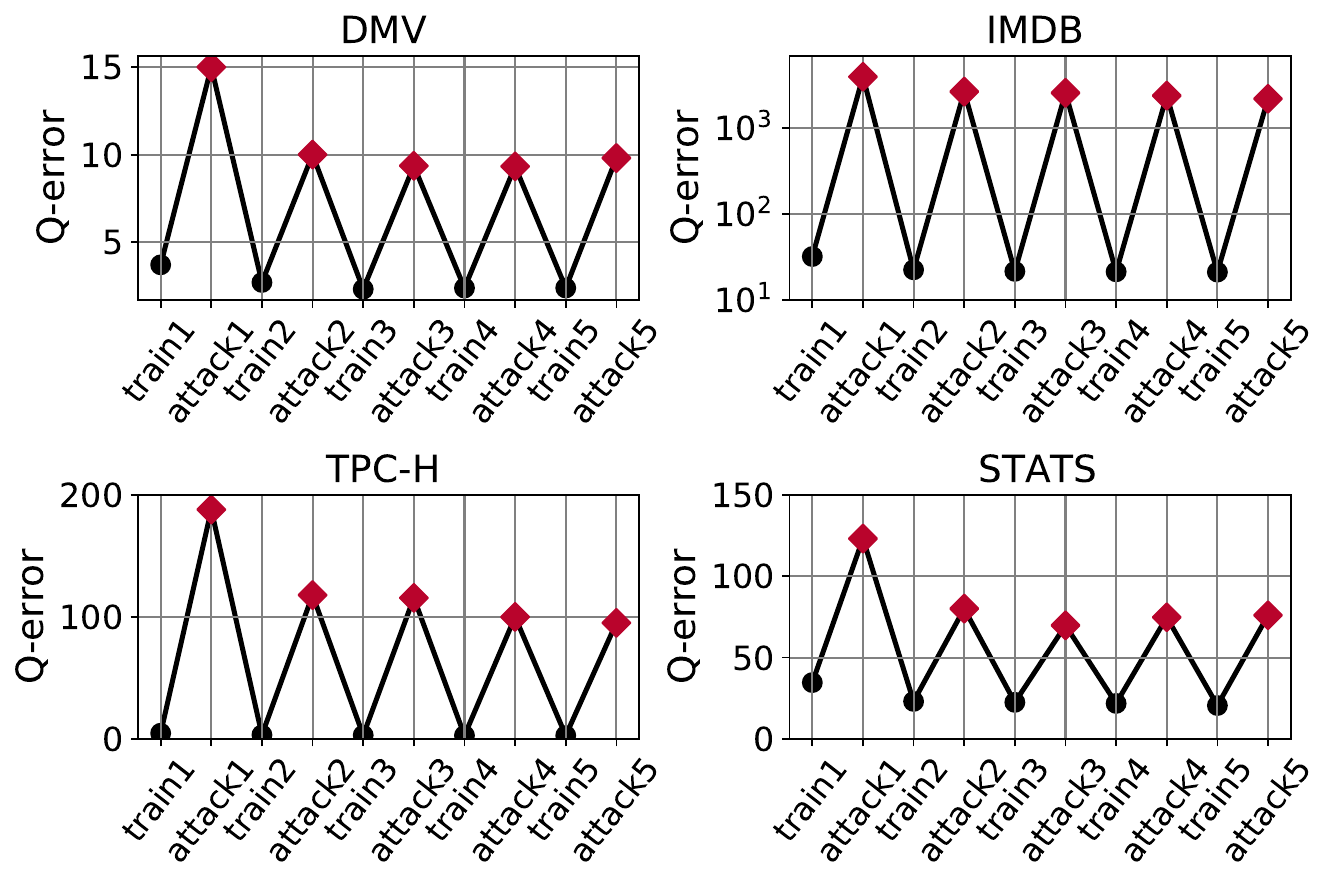}
        \vspace{-.6em}
	\caption{\add{Incrementally training and attacking. The numbers after "train" or "attack" under the x-axis refer to the times of incremental training or attacking.}}
 \vspace{-.5em}
	\label{fig:incremental}
\end{figure}

\subsection{\add{Incrementally Training and Attacking}}  \label{sec:incremental}

\add{We conduct an experiment to explore the effect of \oursys under an incrementally trained CE model. Specifically, the 10000 training queries described in Section~\ref{subsec:exp_setting} are equally divided into five parts on four datasets respectively. Then, we train the \texttt{FCN} model incrementally with the divided training subsets. After each time of incremental training, we evaluated the poisoning effectiveness of \oursys using the testing queries described in Section~\ref{subsec:exp_setting}. The experiment results are shown in Figure~\ref{fig:incremental}, which illustrates that the Q-error of the first time training and attacking are higher than the following ones. That is because, in the early stages of training on a limited number of queries, the CE model had not yet sufficiently captured the relationships between queries and cardinalities. In the following attacks, our system, on average, increases the Q-error of the CE model by an average factor of 22.4$\times$ after each round of incremental CE model training. These experimental results demonstrate the effectiveness and stability of our system.}

\subsection{Effect of the Anomaly Detector} \label{sec:exp_normality}

We determine the abnormality of poisoning queries by comparing its distribution divergence with historical queries. 

As shown in Figure~\ref{fig:effective_of_detector}, we vary the reconstruction error threshold $\epsilon$ from 5\% to 10\%, and compare the effectiveness and normality of poisoning queries of \oursys with and without the anomaly detector on \texttt{DMV} dataset. We note the two methods as \texttt{PACE-Without Detector} and \texttt{PACE-With Detector}. 
We find that \texttt{PACE-With}\\\texttt{Detector} has a 7.6\% decrease in attack effectiveness compared to \texttt{PACE-Without Detector}, but it decreases the abnormality of poisoning query by 72\%. 
That is, using the anomaly detector against the generator does not significantly reduce the poisoning effectiveness of the poisoning queries but ensures the poisoning queries follow similar distribution to the historical workload successfully. 

This is because we are solving a dual-objective optimization problem, which is to increase the poisoning effectiveness of the poisoning queries and at the same time ensure the poisoning queries normality. 
\add{Moreover, the smaller $\epsilon$ is, the less divergence is, but the poisoning effectiveness is worse. We recommend choosing 5\% for $\epsilon$ when using \oursys because it maximizes the ratio of the Q-error of the poisoned model and the divergence between poisoning and history queries.}

\begin{figure}[htbp]
  
	\centering
	\includegraphics[width=0.49\textwidth]{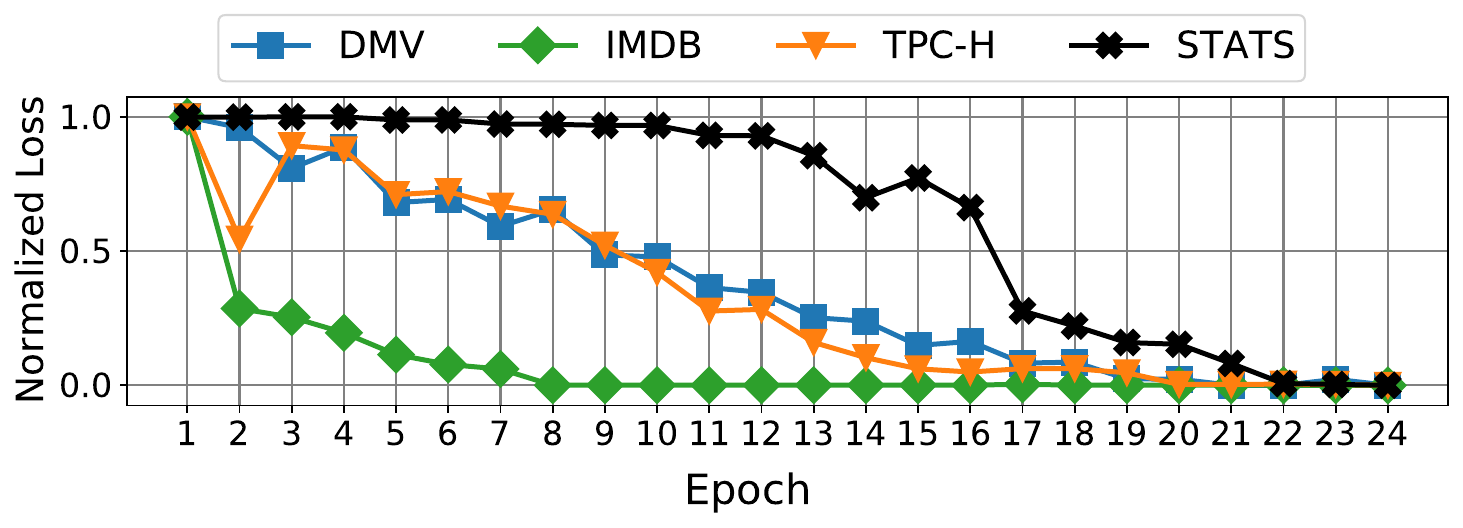}
        \vspace{-.9em}
	\caption{\add{Convergence curve of the optimization objective. The value has been normalized to [0,1].}}
\vspace{-.1em}
	\label{fig:Convergence}
\end{figure}

\subsection{\add{Convergence of the Optimization Objective}} \label{sec:exp_convergence}

\add{In order to verify the feasibility of convergence of the optimization problem in Equation~\ref{equ:opt_obj_}, we report the changing of the loss function value of the optimization objective in the optimization process on \texttt{FCN} on four datasets. The results are shown in Figure~\ref{fig:Convergence}. We can find that despite occasional fluctuations, the overall trend continues to decline and converges ultimately.}
%!TEX root=../main.tex

\section{Conclusion and Future Work} \label{sec:con}

In this work, we study a new problem of poisoning attack on learned cardinality estimation in a black-box setting, and propose a poisoning attack system, \oursys.
We devise a method to replace the black-box model with a white-box surrogate model. Then, we design an algorithm to efficiently train a generator that can craft effective poisoning queries. To ensure the poisoning queries follow a similar distribution to historical workload, we propose an anomaly detector against the generator. Experiments show that \oursys can efficiently and significantly reduce the accuracy of the CE models.

There are two potential directions for further investigation. 

\noindent\add{\textbf{Improve the learned database systems.} There are three ways to improve learned database systems directly based on \oursys. (1) We can train a classifier to detect abnormal queries by using poisoning queries generated by \oursys as training data, and then the classifier can help the learned database systems avoid the attack from poisoning queries. (2) We can test the vulnerability of various cardinality estimation models and recommend a robust one for the learned database systems. (3) As \oursys can be adapted to attack other learned regression models, we can apply it to other learned components~\cite{zhou2021learned, han2022dynamic, zhou2023grep, sun2023learned, li2022machine, li2021ai, han2021autonomous, huang2023acr} and improve their security.}

\noindent\add{\textbf{Extend to a budget-constrained setting.} Typically, the attacker has a limited budget for an attack. Thus there should be a budget that constrains the number of the poisoning queries. One possible  approach is to adjust the corresponding parameter of \oursys to generate fewer queries. But in order to maximize the poisoning effect of a limited number of poisoning queries, we should design a penalty function, which represents a penalty on the optimization objective when the constraints are not met. This approach allows the solution of the unconstrained problem to converge towards the solution of the constrained problem.}

\begin{acks}
This paper was supported by National Key R\&D Program of China (2023YFB4503600), NSF of China (61925205, 62232009, 62102215), Science and Technology Research and Development Plan of China Railway (K2022S005), Huawei, CCF-Huawei Populus Grove Challenge Fund (CCF-HuaweiDBC202309), TAL education, and Beijing National Research Center for Information Science and Technology (BNRist).
\end{acks}

\bibliographystyle{ACM-Reference-Format}
\bibliography{main}

\end{document}